\documentclass[useAMS,usenatbib]{mn2e}
\usepackage{graphicx}

\title[Two-component anisotropic polytropes]{Analytical families of two-component anisotropic polytropes and their relativistic extensions}
\author[P. H. Nguyen and M. Lingam]{Phuc H. Nguyen$^{1}$\thanks{E-mail:
phn229@physics.utexas.edu} and Manasvi Lingam$^{2}$\thanks{E-mail:
manasvi@physics.utexas.edu}\\
$^{1}$Center for Relativity and Texas Cosmology Center, The University of Texas, Austin, TX 78712, USA\\
$^{2}$Institute for Fusion Studies, The University of Texas, Austin, TX 78712, USA}
\begin{document}

\date{Accepted 2013 September 10. Received 2013 August 8;}

\pagerange{\pageref{firstpage}--\pageref{lastpage}} \pubyear{2013}

\maketitle

\label{firstpage}

\begin{abstract}
In this paper, we study a family of two-component anisotropic polytropes which model a wide range of spherically symmetric astrophysical systems such as early-type baryonic galaxies. This family is found to contain a large class of models such as the hypervirial family (which satisfy the virial theorem locally), the Plummer and Hernquist models and Navarro-Frenk-White (NFW)-like models. The potential--density pair for these models are derived, as well as their velocity dispersions and anisotropy parameters. The projected quantities are computed and found to reduce to analytical expressions in some cases. The first section of this paper presents an extension of the two-term anisotropic polytropes to encompass a very wide range of potential--density pairs. In the next section, we present the general relativistic extension of the potential--density pair, and calculate the stress--energy tensor, the relativistic anisotropy parameter, the velocity of circular orbits and the angular momentum. Remarkably, for the case of the hypervirial family, the relativistic pressure in the Newtonian limit and the relativistic anisotropy parameter are found to coincide with the corresponding Newtonian expressions. The weak, dominant and strong energy conditions are found to be satisfied only for a certain range of the free parameters. We show that the relativistic hypervirial family also has a finite total mass like its Newtonian counterpart. In the first appendix, a relativistic extension of a different hypervirial family of models is studied, and the relativistic anisotropy parameter is found to coincide with the Newtonian one. Finally, we present a family of models obtained from our distribution function that are similar to the Ossipkov--Merritt models; by computing their anisotropy parameters, we show that they model systems with isotropic cores and radially anisotropic exteriors.
\end{abstract}


\section{Introduction}
The simplest class of three-dimensional models that we can study analytically usually incorporate spherical symmetry. Although there are not many astrophysical systems that possess such an idealized symmetry, spherical models are still very useful in studying globular clusters, galactic bulges and dark matter haloes. In order to model such systems in a self-consistent manner, the Vlasov--Poisson system must be solved, which necessitates a knowledge of the distribution function $f$. A commonly used approach (known as the ``$\rho$ to $f$'' approach) starts with the potential--density pair, which is inverted to obtain the distribution function \citep{b2x,c21,b252,y123,bt87,c22,b2z01}. However, a limitation of such an approach stems from the fact that the distribution functions thus obtained are highly cumbersome and complex, and do not lend themselves to analytical calculations. 

In this paper, we follow the reverse approach (the ``$f$ to $\rho$ approach'') and start with a simple two-component, two-parameter family of distribution functions, which is found to give rise to a potential--density pair that includes the hypervirial models which were discussed in \citet{b000}, which is henceforth referred to as Paper I, and \citet{gf21}, which in turn includes the Plummer and Hernquist models as specific cases. Our family also includes a specific case of the Ossipkov--Merrit models and an NFW-like profile. The potential--density pair that is obtained was recently discussed in \citet{x22}, where distribution functions of the isotropic or Opsikov--Merritt form for this potential--density pair were also derived using well-known prescriptions. These distribution functions are more complicated than the one presented in this paper, but they are designed to give rise to an isotropic core and an isotropic or radially anisotropic outer region, a feature that our distribution function does exhibit for some special cases, as outlined in Appendix B.

Our family of distribution functions depend on both energy and the norm of angular momentum (which are constants of motion in the spherically symmetric setting), and therefore belong to the very general class of two-integral distribution functions \citep{b123}. There exists a rather large literature on prescriptions to yield two-integral distribution functions. In the axisymmetric setting, \cite{b21} noticed that whenever the density profile can be written as a power of the radial coordinate times a power of the potential (up to a constant of proportionality), or a sum of such terms, then we can write down a distribution function consisting of powers of angular momentum and energy. This procedure was then applied by different authors \citep{m63,pt70,k76,jf85,dj86,r88,e93,e94,b2z01b,b2z01c} to study a large class of galactic models, and extended by \cite{b423} and \cite{b2z01} to more general functional dependences of the density on the radial coordinate and the potential. The approach devised in this paper is similar in spirit to the ones mentioned.

While the distribution function used in Paper I has the generalized polytrope form, i.e. a power of angular momentum times a power of the relative energy, the model presented in this paper generalizes this approach by constructing a superposition of two such terms. Importantly, this generalization allows the anisotropy parameter to be non-constant (unlike in Paper I). Interestingly, the family of models presented in Paper I was also derived by \citet{gf21}, who also generalized the model to a two-term distribution function but in a different way from the one presented here (see \citet{d29}).

The possible density profiles can model a wide range of astrophysical systems inclusive of both compact objects and extended systems, and include configurations with a central cusp, with a finite central density, as well as shell-like configurations. The family of spherical models presented in this paper are useful in modelling astrophysical cusps such as the centres of massive early-type galaxies and dwarf spheroidal galaxies. In addition to cusps, shells model a wide range of astrophysical phenomena such as supernovae and gravitational collapse. The family of models presented in this paper allows one to study a wide range of Newtonian and relativistic shells.

In the second half of the paper, we will generalize this potential--density pair to general relativity. It is important to note that there are several spherically symmetric astrophysical systems such as galactic nuclei and certain star clusters where relativistic effects are expected to have a significant impact. Such effects are expected to be particularly significant in rogue nuclei that have become more massive (relativistic) following their severance from the galaxy, and there is sufficient observational evidence that appears to validate such a conclusion \citep{b23}.

The approach that we will take will be to construct an anisotropic analogue of a static spherically symmetric perfect fluid (SSSPF) solution to Einstein's field equations using an approach that was used to derive a relativistic version of the Plummer model. This approach was then successfully applied to other Newtonian potentials by numerous authors, in particular \citet{b94,b95,b96}. It is important to note that we shall not work with distribution functions in this section. However, one can find self-consistent treatments of the post-Newtonian limit in the literature, such as \citet{xf21,g21}, \citet{e21}, \citet{g12}, \citet{h412} and Paper I.

The organization of the paper is as follows. In Section 2, we present the two-term, two-parameter family of distribution functions and the corresponding potential--density pairs that arise from them. Here, we compute the velocity dispersions, anisotropy parameters and projected quantities along with a brief discussion of the hypervirial family developed by \citet{gf21}. In Section 3, we present an outline of a generalized version of the anisotropic polytropes. In Section 4, we present the analogue of the SSSPF approach that is used to compute the components of the stress--energy tensor.  The energy conditions for stability, the relativistic pressures, the anisotropy parameters, the properties of circular orbits and an upper bound for the relativistic hypervirial family are investigated. Appendix A includes a brief discussion of the two-term hypervirial family of \citet{d29}. Lastly, a family of models similar to the Ossipkov--Merritt models is developed from our distribution function, and their anisotropy parameters are computed in Appendix B. 

\section[]{Derivation of the two-term distribution function}
We start with a distribution function of the form
\begin{equation}\label{EQ1}
F(\mathcal{E},L) = A L^{2\alpha} \mathcal{E}^{\beta} + B L^{2\gamma} \mathcal{E}^{\delta},
\end{equation}
where $\mathcal{E}=\phi_{\star}-E$ is the relative energy (for infinite systems, studied in this paper, we note that $\phi_\star =0$) and $L$ is the norm of angular momentum. It is implicitly understood that the distribution function vanishes when the right-hand side of the above expression is negative. Moreover, if $A$ and $B$ are both non-negative, this is equivalent to the condition that $F$ is given by equation (\ref{EQ1}) for $\mathcal{E} \geq 0$ and that $F=0$ for $\mathcal{E} \leq 0$.

This shall be the case that we shall study throughout the paper. If either $A$ or $B$ vanishes, then the distribution function reduces to the generalized polytropic ansatz extensively studied in the literature. To solve for the gravitational potential $\phi$, we use the Poisson equation:
\begin{equation}
\nabla^2 \phi = 4 \pi G \rho,
\end{equation}
where the density can be found by integrating the distribution function over all velocities:
\begin{equation}
\rho = \int f d^3 \bf{v}.
\end{equation}
By substituting the expression for the distribution function and carrying out the integral in spherical coordinates in velocity space $(v,\eta,\zeta)$, we have
\begin{eqnarray}\label{rho}
\rho &=& A \int_0^{2\pi}\int_0^{\pi}\int_0^{v_e} L^{2\alpha} {\mathcal{E}}^{\beta} v^2 dv \sin \eta d\eta d\zeta  \\ \nonumber 
&+& B \int_0^{2\pi}\int_0^{\pi}\int_0^{v_e} L^{2\gamma} {\mathcal{E}}^{\delta} v^2 dv \sin \eta d\eta d\zeta,
\end{eqnarray}
where the escape velocity $v_{e}$ is found by solving the equation $\mathcal{E}=0$, and we introduce the notation $\mathrm{\Phi} = \phi - \phi_{\star}$. For the models studied in this paper, $\phi_\star = 0$, as was noted earlier. The escape velocity is given by
\begin{equation}
v_e = \sqrt{-2\Phi}.
\end{equation}
The square of the angular momentum is given in spherical coordinates in velocity space by
\begin{equation}
L^2 = r^2 v^2 \sin^{2} \eta.
\end{equation}
Upon substituting these into equation (\ref{rho}), the density is
\begin{eqnarray}
\rho &=& 2^{\alpha+\frac{3}{2}}\pi^{3/2}A\Gamma(\alpha,\beta) r^{2\alpha}(-\mathrm{\Phi})^{\alpha+\beta+\frac{3}{2}} \\ \nonumber 
&+& 2^{\gamma+\frac{3}{2}}\pi^{3/2}B \Gamma(\gamma,\delta) r^{2\gamma}(-\mathrm{\Phi})^{\gamma+\delta+\frac{3}{2}},
\end{eqnarray}
provided $\alpha,\gamma, \beta, \delta > -1$ (otherwise the integrals in equation (\ref{rho}) diverge and the mass density is everywhere infinite), and we define the function $\Gamma(a,b)$ to be:
\begin{equation}
\Gamma\left(a,b\right) = \frac{\Gamma{(1+a)}\Gamma{(1+b)}}{\Gamma\left(a+b+\frac{5}{2}\right)}.
\end{equation}
The Poisson equation is then:
\begin{equation}
\nabla^2 \Phi = C_1 r^{2\alpha} (-\mathrm{\Phi})^{\alpha+\beta+\frac{3}{2}} + C_2 r^{2\gamma} (-\mathrm{\Phi})^{\gamma+\delta+\frac{3}{2}},
\end{equation}
where
\begin{equation}
C_{1} = 2^{\alpha+\frac{7}{2}}\pi^{5/2} G A \Gamma(\alpha,\beta),
\end{equation}
\begin{equation}
C_{2} = 2^{\gamma+\frac{7}{2}}\pi^{5/2} G B \Gamma(\gamma,\delta).
\end{equation}
To motivate our approach, note that the Hernquist and the Plummer models have a potential which takes on the form
\begin{equation}
\Phi \propto - \frac{1}{\left(1+\left(\frac{r}{a}\right)^m\right)^{1/m}},
\end{equation}
with $m=1$ for the Hernquist model and $m=2$ for the Plummer model. In Paper I, this family of models was obtained from a one-term distribution model. The above potential corresponds to the hypervirial family that was derived by \citet{gf21}. Now, we wish to generalize this method and incorporate a second parameter into the above expression. To do so, we write the four powers $\alpha$, $\beta$, $\gamma$ and $\delta$ as functions of two free parameters $m$ and $n$ as follows:
\begin{equation}
\alpha(m,n) = m - 1,
\end{equation}
\begin{equation}
\beta(m,n) = \frac{2}{n} + \frac{1}{2} - m,
\end{equation}
\begin{equation}
\gamma(m,n) = \frac{m}{2} - 1,
\end{equation}
\begin{equation}
\delta(m,n) = \frac{2}{n} + \frac{1}{2} - \frac{m}{2}.
\end{equation}
The inequalities for $\alpha$, $\beta$, $\gamma$ and $\delta$ translate to the following inequalities in $m$ and $n$:
\begin{equation}\label{ineq1}
m > 0,
\end{equation}
\begin{equation}\label{ineq2}
\frac{2}{n} > m - \frac{3}{2}.
\end{equation}
Moreover, we will impose a constraint on the two constants $A$ and $B$ as follows (the reason why will become apparent soon):
\begin{equation}
\frac{C_1}{C_2} = \frac{1-nm}{m+1} a^{-m}.
\end{equation}
We note that $a$ is the length-scale introduced earlier in the hypervirial potential and this also ensures that the ratio of $C_1$ and $C_2$ is dimensionally correct.
The distribution function in terms of $m$ and $n$ then takes the form:
\begin{eqnarray} \label{DF0}
f &=& B \bigg[ L^{m-2} \mathcal{E}^{\frac{2}{n} + \frac{1}{2} - \frac{m}{2}} + 2^{-m/2} \frac{\Gamma\left(\frac{m}{2}\right)\Gamma\left(\frac{2}{n}+\frac{3-m}{2}\right)}{\Gamma\left(m\right)\Gamma\left(\frac{2}{n}+\frac{3-2m}{2}\right)} \nonumber  \\
  &\times& a^{-m}\left(\frac{1-nm}{m+1}\right) L^{2m-2} \mathcal{E}^{ \frac{2}{n} + \frac{1}{2} - m} \bigg],
\end{eqnarray}
for some overall constant $B$. Since our distribution function has a built-in condition that $A$ and $B$ are both non-negative, this amounts to the condition that $mn \leq 1$. For the particular case $mn=1$, we recover the distribution function discussed in Paper I as well as \cite{gf21}:
\begin{equation}\label{DFhypervirial}
f = B \bigg[ L^{m-2} \mathcal{E}^{ \frac{3m+1}{2} } \bigg].
\end{equation}
For convenience, we relegate the discussion of the hypervirial family to Section \ref{sechypervirial} and come back to the more general family for the remainder of this section. The Poisson equation in terms of $m$ and $n$ takes the form:
\begin{equation}
\nabla^{2}\Phi = C_{2}\left[r^{m-2} + \left(\frac{1-nm}{m+1}\right)a^{-m}r^{2m-2} \right](-\Phi)^{1+\frac{2}{n}}
\end{equation}
and admits the following particular solution:
\begin{equation}
\Phi = -\frac{\Phi_{0}}{\left(1+\left(\frac{r}{a}\right)^m\right)^n},
\end{equation}
where the constant $\Phi_0$ is taken to be positive, and can be written as a function of $B$. As pointed out by \citet{x22}, this family of potential encompasses a very broad class of power-law-like models found in the literature. The distribution functions derived by \citet{x22}, however, are different from the one presented in this paper. Also, since the potential must be bounded as $r \rightarrow \infty$, we impose the extra constraint on the range of $n$:
\begin{equation}
n > 0.
\end{equation}
Combining the above inequality with inequalities (\ref{ineq1}) and (\ref{ineq2}) yields the allowed range of the parameters $m$ and $n$ as follows: if $m \leq \frac{3}{2}$, $n$ can take any value from 0 to $\infty$, and if $m > \frac{3}{2}$, there is an upper bound for $n$ given by
\begin{equation}
n < \frac{4}{2m-3}.
\end{equation}
Next, the density is found to be
\begin{equation}
\rho(r) = \frac{mn{a^{mn}}\Phi_0}{4\pi{G}}\frac{(1+m)a^m+(1-mn)r^m}{r^{2-m}\left(a^m+r^m\right)^{n+2}}.
\end{equation}
As the properties of this potential--density pair have been extensively investigated in \citet{x22}, we will restrict ourselves to a discussion of the most important features. Note that since we specialized to the case $1-mn \geq 0$, the mass density is everywhere non-negative.

For the case where $mn \geq 1$, on the other hand, the two terms in the distribution function are of opposite sign. We will not solve the Poisson equation for models of this type, but we will explore a few of their properties in Appendix \ref{appOssipkovMerritt}. 

One could obtain the potential--density pair for the values of $m$ and $n$ that fall under this category, i.e. which satisfy $mn \geq 1$. But, it must be noted that the above potential--density pair is $not$ the same as the one obtained by solving the Vlasov--Poisson system. One must impose a specific truncation radius $r_\star$ for these models, where $\rho$ changes sign. The value of $r_\star$ is
\begin{equation}\label{cutoffr}
r_\star = a \left(\frac{1+m}{mn-1}\right)^{1/m}.
\end{equation}
Also, it is of interest to investigate the small distance and large distance limits of the density profile. The large $r$ limit gives the power--law dependence:
\begin{equation}
\rho{(r)} \propto r^{-2-mn},
\end{equation}
for $mn \neq 1$. We note that in this case, with $mn < 1$, the density falls off more slowly than $r^{-3}$, and hence, the total mass diverges. We can, of course, truncate the system at an arbitrary radius if we wish to have a finite total mass. For $mn = 1$, the outer cusp is:
\begin{equation}
\rho{(r)} \propto r^{-3-m}.
\end{equation}
In this case, $\rho$ falls off faster than $r^{-3}$ and the total mass is finite. Next, we investigate the behaviour of $\rho$ at small $r$. We find that the inner cusp is qualitatively different depending on whether $m < 2$, $m=2$ or $m>2$. For $m<2$, the density diverges as $r \rightarrow 0$ and we have
\begin{equation}
\rho{(r)} \propto r^{m-2}.
\end{equation}
For $m=2$, the value of the mass density at $r=0$ is finite and non-zero, while for $m>2$, the density vanishes at the centre, has a maximum near $r=a$ and we have a shell-like configuration.

As an application, we look for models which exhibit cosmologically important cusps. In particular, we can use the inner cusp to infer the nature of dark matter particles \citep*[see, for example][]{b569,b00}. An inner cusp $\rho \propto r^{-1}$, for example, is obtained by choosing $m=1$. This inner cusp is a feature of the NFW profile, which also has outer cusp $\rho \propto r^{-3}$. Even though no model in our family exhibits these very asymptotic behaviours, we can obtain excellent approximations to the NFW profile by choosing $m=1$ and $n \approx 1$ (but $n \neq 1$). The particular choice $m=1$ and $n=\frac{1}{2}$ has already been considered in the literature \citep{b14}. In this case, the potential--density pair is given by
\begin{equation}
\Phi = -\frac{\Phi_{0}}{\sqrt{1+\left(\frac{r}{a}\right)}},
\end{equation}
\begin{equation}
\rho(r) = \frac{\Phi_{0}}{8\pi G a^{2}} \frac{2+\frac{1}{2}\frac{r}{a}}{\frac{r}{a}\left(1+\frac{r}{a}\right)^{5/2}}.
\end{equation}
Other inner cusps of cosmological significance are $\rho \propto r^{-4/3}$, which is obtained for $m=\frac{2}{3}$ and $\rho \propto r^{-3/2}$, which is obtained for $m=\frac{1}{2}$ (see \citet{b569} and \citet{g681}). Also, the Bahcall-Wolf cusp $\rho \propto r^{-7/4}$ is obtained for $m = \frac{1}{4}$ \citep{bw76}. We note that the inner cusp is determined only by $m$ (and not $n$); hence, for each of the above values of the inner cusp we have a subfamily labelled by $n$ all of whose members have the given cusp.

\subsection{Velocity dispersion, pressure and anisotropy parameter}
\begin{figure*}
$$
\begin{array}{ccc}
 \includegraphics[width=5.28cm]{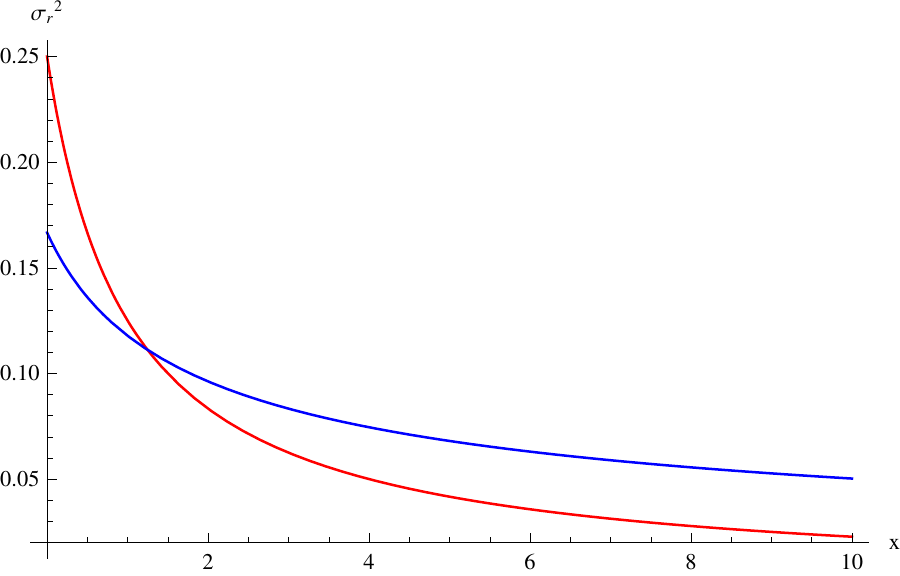} & \includegraphics[width=5.28cm]{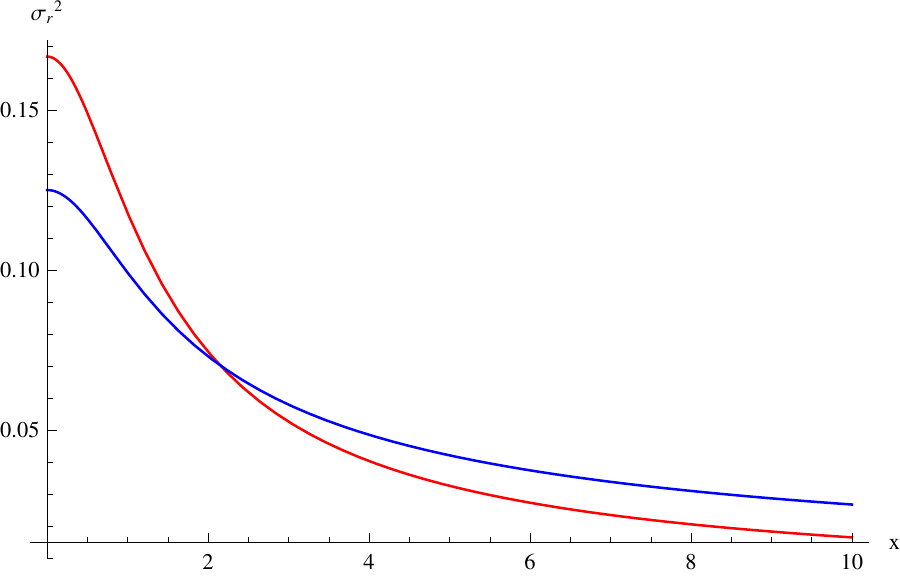} & \includegraphics[width=5.28cm]{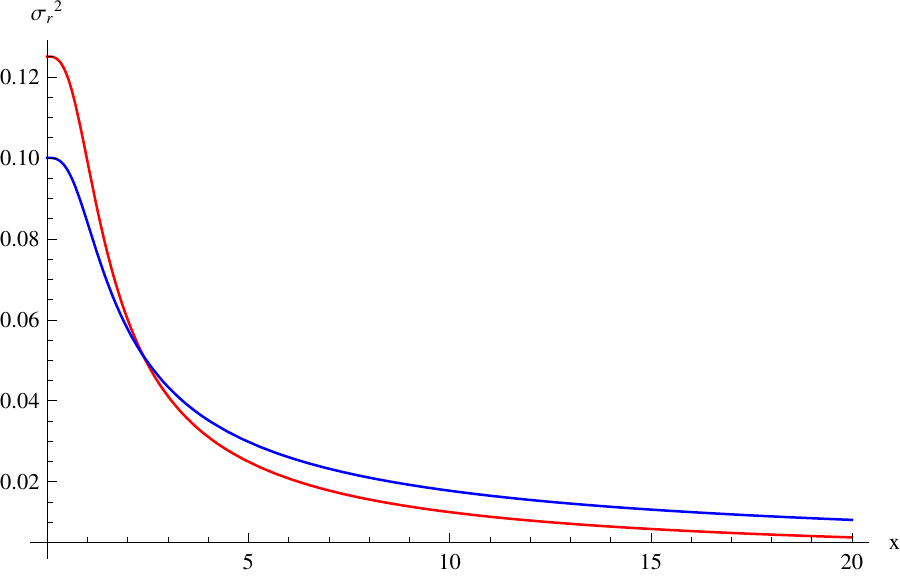}\\
 \quad\quad(a) & \quad\quad(c) & \quad\quad(e)\\
  \includegraphics[width=5.28cm]{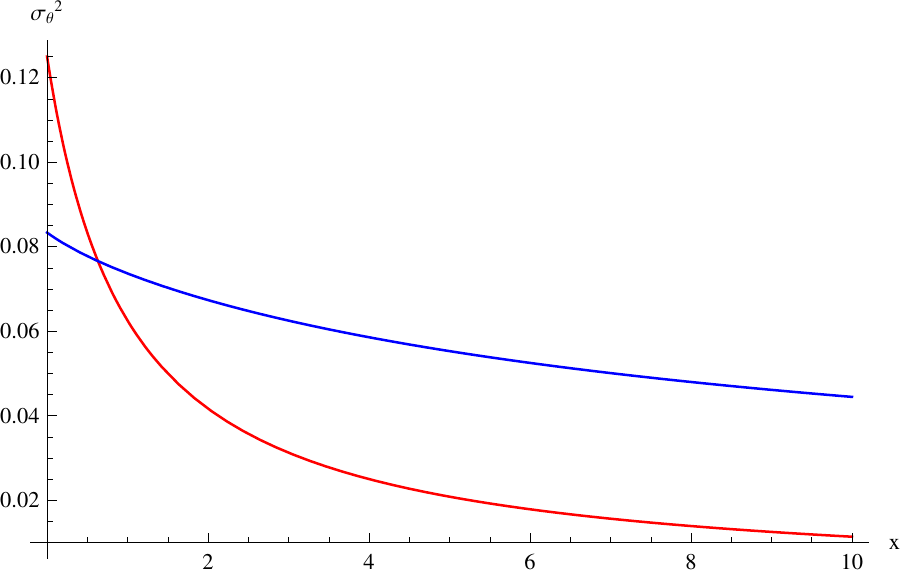} & \includegraphics[width=5.28cm]{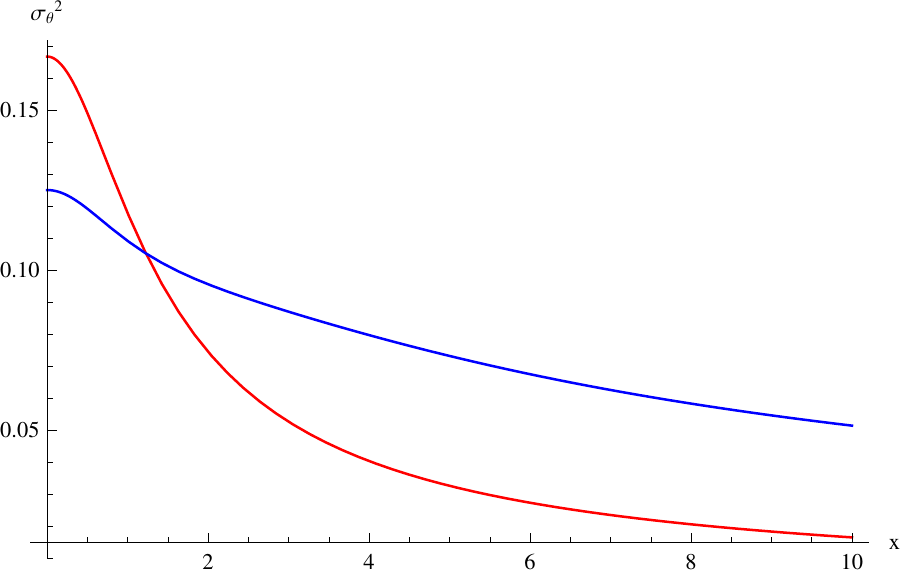} & \includegraphics[width=5.28cm]{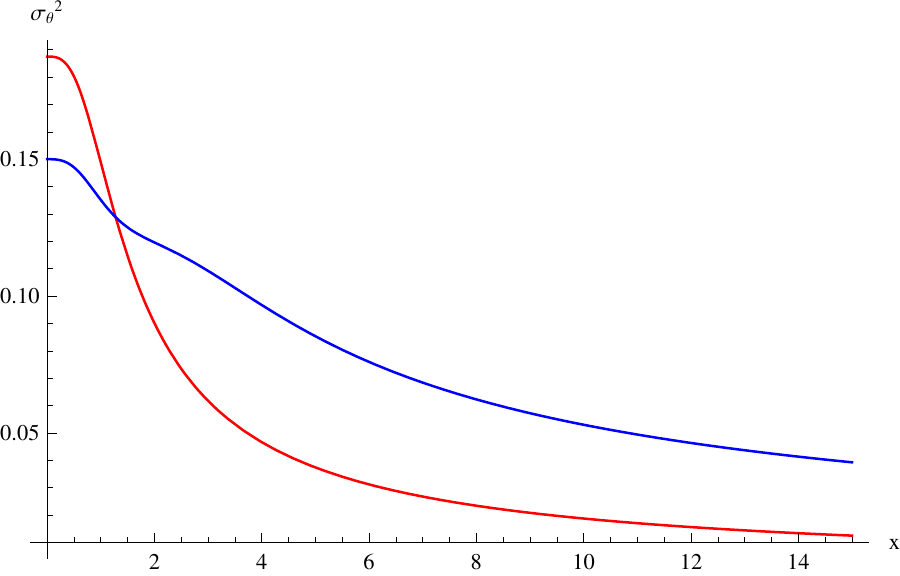}\\
 \quad\quad(b) & \quad\quad(d) & \quad\quad(f)\\
\end{array}
$$
\caption{(color figures online) The $x$-axis is expressed in units of $x=\frac{r}{a}$, and the $y$-axis is chosen to be $\sigma_r^2$ for (a), (c) and (e), while it is $\sigma_\theta^2$ for (b), (d) and (f). Figures (a) and (b) show plots of the dispersion curves for Hernquist (red) and NFW-like profile (blue) and both constitute cases of cusp-like profiles. Figures (c) and (d) show plots for the Plummer (red) and a case with $m=2$, $n=\frac{1}{3}$ (blue). These two profiles have a finite central density. The Figures (e) and (f) show plots for $m=3;\,n=\frac{1}{3}$ (red) and $m=3;\,n=\frac{1}{4}$ (blue) profiles which are shell-like. Here, we have chosen the value of $\Phi_0=1$ for the sake of simplicity.}
\label{figdisp1}
\end{figure*}

\begin{figure*}
$$
\begin{array}{ccc}
 \includegraphics[width=5.2cm]{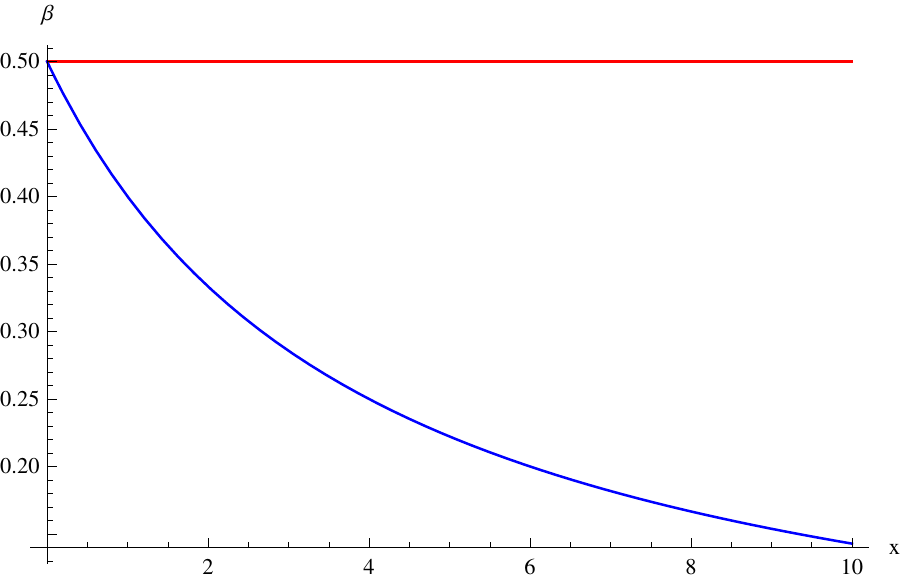} & \includegraphics[width=5.2cm]{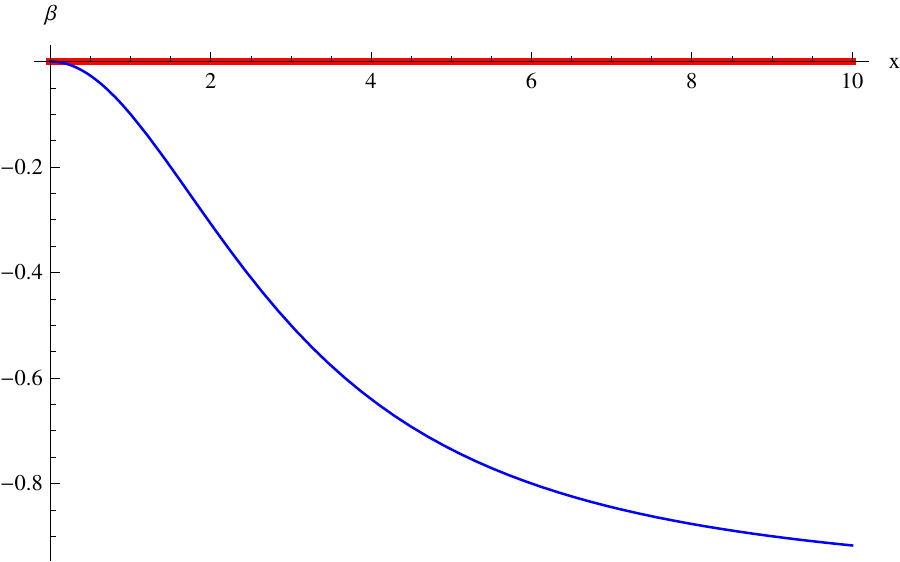} & \includegraphics[width=5.2cm]{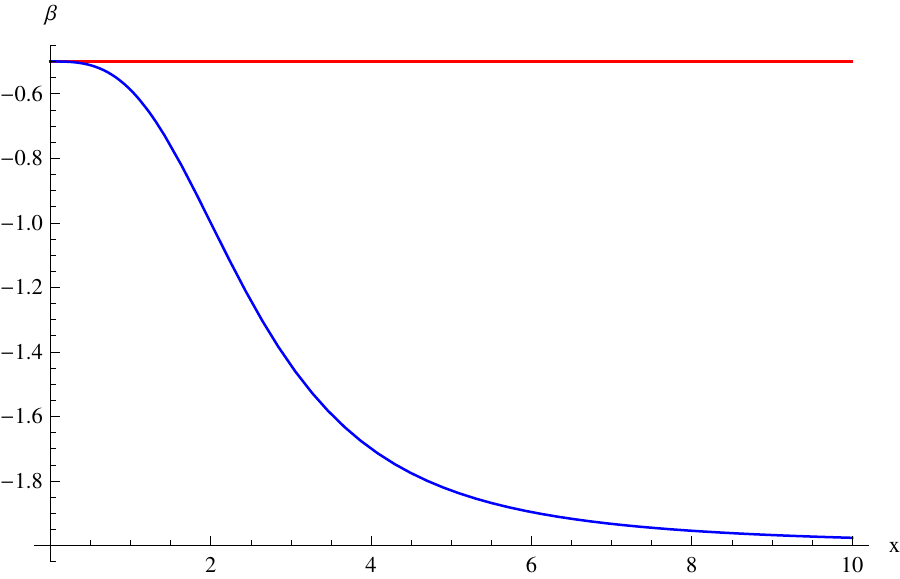}\\
 \quad\quad(a) & \quad\quad(b) & \quad\quad(c)\\
\end{array}
$$
\caption{(color figures online) The $x$-axis is expressed in terms of $x=\frac{r}{a}$ while the $y$-axis is a measure of the anisotropy parameter $\beta$. Figure (a) plots $\beta$ for the Hernquist (red) and NFW-like (blue) profiles; figure (b) for the Plummer (red) model and a case with $m=2$, $n=\frac{1}{3}$ (blue); figure (c) shows a plot for the $m=3;\,n=\frac{1}{3}$ (red) and $m=3;\,n=\frac{1}{4}$ (blue) profiles. Note that the three red lines are constant, and this arises from the fact that they are all members of the hypervirial family, which possess a constant anisotropy parameter.}
\label{figbeta}
\end{figure*}

In this section, we compute the velocity dispersion tensor and associated quantities, such as the pressure tensor or the anisotropy parameter. Since the velocity dispersion is specific to the distribution function, these results do not have any overlap with those presented in \citet{x22}, which computes the velocity dispersion for the same potential--density pair but using different distribution functions. We use the following relations \citep{dj87}:
\begin{equation}
\sigma_r^2(r) \equiv \langle{v^2_r}\rangle = -\frac{1}{\rho(r,\phi)} \int_0^\phi{\rho(r,\phi')}d\phi',
\end{equation}
\begin{equation}
\sigma_\phi^2(r) \equiv \langle{v^2_\phi}\rangle = -\frac{1}{\rho(r,\phi)} \int_0^\phi{\partial_{r^2}\left[r^2\rho(r,\phi')\right]}d\phi',
\end{equation}
where $\rho{(r,\phi)}$ is found by the fundamental integral relation between the distribution function and the mass density, and is found to be:
\begin{equation}
\rho(r,\phi) \propto \frac{(1+m)a^m+(1-mn)r^m}{r^{2-m}} (-\phi)^{1+\frac{2}{n}}.
\end{equation}
Thus, we obtain
\begin{equation}\label{sigmar}
\sigma_r^2(r) = -\frac{1}{2}\frac{n}{n+1}\phi,
\end{equation}
\begin{equation}\label{sigmat}
\sigma_\phi^2(r) = - \frac{m}{4}\frac{n}{n+1}\left[{1+\frac{1}{1+\left({\frac{m+1}{1-mn}}\right)\left({\frac{a}{r}}\right)^m}}\right] \phi.
\end{equation}
The radial and tangential velocity dispersions are plotted for several models in Fig. \ref{figdisp1}. It is important to note that the velocity dispersion must always be strictly positive. Upon imposing this constraint, the radial and tangential components of the velocity dispersion are found to be positive for all values of $r$, provided that the condition $mn \leq 1$ is satisfied. 
The radial pressure is given by
\begin{equation}
P_{\mathrm{r}} = \rho(r) \sigma_r^2(r),
\end{equation}
\begin{equation}\label{Pr}
P_{\mathrm{r}} = \frac{mn^2{\Phi_0}^2}{8{\pi}G(n+1)r^2}\left(\frac{r}{a}\right)^m \frac{(1+m)+(1-mn)\left(\frac{r}{a}\right)^m}{\left(1+\left(\frac{r}{a}\right)^m\right)^{2n+2}}.
\end{equation}
The tangential components of the pressure are given by
\begin{equation}\label{Pt}
P_\theta = P_\phi = \rho(r) \sigma_\phi^2(r) = \frac{m}{2} \left[{1+\frac{1}{1+\left({\frac{m+1}{1-mn}}\right)\left({\frac{a}{r}}\right)^m}}\right] P_{\mathrm{r}}.
\end{equation}
The anisotropy parameter is defined to be
\begin{equation}\label{anisotropyparameter}
\beta = 1 - \frac{\sigma_\phi^2(r)}{\sigma_r^2(r)} = 1 - \frac{m}{2} \left[{1+\frac{1}{1+\left({\frac{m+1}{1-mn}}\right)\left({\frac{a}{r}}\right)^m}}\right].
\end{equation}
The anisotropy parameter is plotted for different models in Fig. \ref{figbeta}. We note that, for the special case of $mn=1$, the anisotropy parameter becomes a constant, i.e. it does not depend on the value of $r$. We recover the same expression obtained in Paper I, given by
\begin{equation}\label{anisotropyparameterhypervirial}
\beta = 1 - \frac{m}{2}.
\end{equation}
Next, consider the case $mn \leq 1$. In this case, the values of the anisotropy parameter at the centre and at infinity are:
\begin{equation}
\beta{(0)} = 1 - \frac{m}{2},
\end{equation}
\begin{equation}
\beta{(\infty)} = 1 - m.
\end{equation}
We now have three subcases: when $m < 1$, when $1 \leq m \leq 2$ and when $2 < m$. In the first subcase ($m < 1$), the anisotropy parameter is everywhere positive. In the third subcase ($m > 2$), the anisotropy parameter is everywhere negative. Finally, for $1 \leq m \leq 2$, the anisotropy is positive in the inner region and negative in the outer region. The value of $r$ where the anisotropy parameter changes sign is given by
\begin{equation}
r_0 = a \left[{\frac{(2-m)(m+1)}{2(m-1)(1-mn)}}\right]^{1/m}.
\end{equation}
The sign of the anisotropy parameter can be physically interpreted as follows: when $\beta < 0$, then nearly circular orbits are preferred; when $\beta > 0$, then nearly radial orbits are preferred, and when $\beta=0$, the two types of orbits are equally probable.

Thus, for $0 \leq m \leq 1$, the model is radially anisotropic, with a stronger anisotropy in the core than in the outer region. For $1 \leq m \leq 2$, the core is slightly radially anisotropic while the outer region is tangentially anisotropic. For $m > 2$ (shell-like configurations), both the core and the outer region are tangentially anisotropic, with a stronger bias in the outer region. As far as dark matter haloes are concerned, $N$-body simulations suggest an isotropic core and a radially biased outer region (\cite{hm06}), and our models are not well--suited for this purpose. Nevertheless, we find applications of a non-isotropic galactic core discussed in \cite{bvh07} and \citet{d212}, which attempts to construct dynamical models corresponding to an arbitrary anisotropy profile. Also, a tangentially anisotropic outer region is discussed in \cite{ae06}, according to whom such a profile is useful in modelling subhaloes or satellite galaxies. In Appendix \ref{appOssipkovMerritt}, we compute the anisotropy parameter for a few cases with $mn \geq 1$ (the anisotropy parameter can be computed even though we do not solve the Poisson equation), and we show that for this subclass of models, we can have an isotropic core and a radially anisotropic outer region which is ideal for dark matter haloes.

\subsection{Projected quantities}
\begin{figure}
 \quad\quad\quad \includegraphics[width=5.68cm]{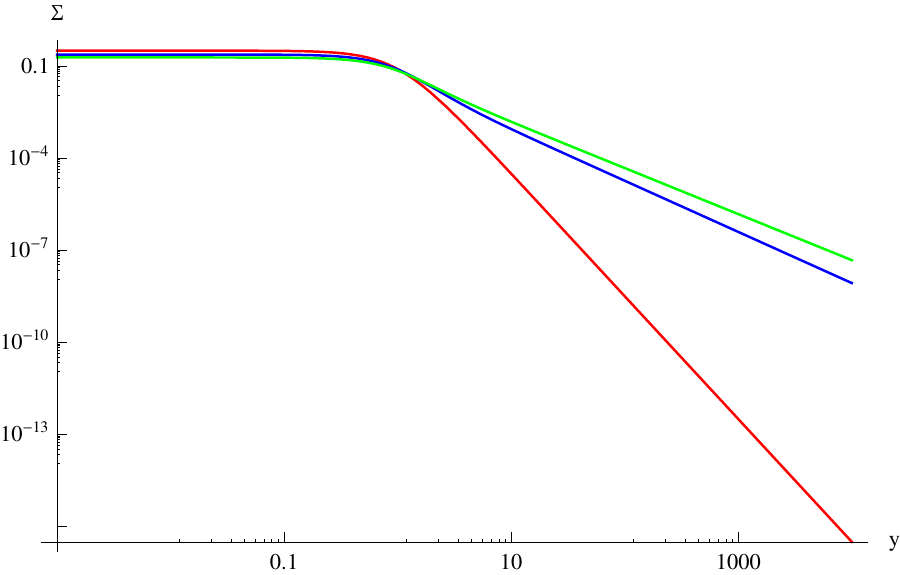} \\
\caption{(color figures online) The figure plots $y=\frac{R}{a}$ on the x-axis and the dimensionless surface density on the y-axis given by $\tilde{\Sigma} = \Sigma \frac{Ga}{\Phi_0}$. The three cases plotted are for the case with $m=2$, but with differing values of $n$. The red line represents the Plummer model, the blue line represents the model with $n=\frac{1}{3}$ and the green one represents the model with $n=\frac{1}{4}$. Note that this is a log-log plot. }
\label{figdisp2}
\end{figure}

Even though the distribution function is not directly measurable, there are some quantities closely related to the distribution function that are accessible by photometric and kinematic observations, namely the projected density and the light-of-sight velocity dispersion. In this section, we will compute the surface density for a few particular cases of our family (the line-of-sight velocity dispersion is usually not analytically computable). This quantity is given by:
\begin{equation}
\Sigma = 2\int_R^\infty \frac{{\rho}rdr}{\sqrt{r^2-R^2}}.
\end{equation}
As pointed by \citet{gf21}, this integral cannot be done analytically even for the particular case of the hypervirial family, except for the Plummer and Hernquist. The central value of the surface density, however, can be computed for general $m$ and $n$. It diverges for $m \leq 1$, and is finite for $m>1$. In the latter case, it is given by:
\begin{equation}
\Sigma{(R=0)} = -\frac{n\Phi_{0}}{m\pi G a} \frac{\Gamma{(-\frac{1}{m})}\Gamma{(n+\frac{1}{m})}}{\Gamma{(1+n)}}.
\end{equation}
Notice that since $m>1$, the factor $\Gamma{(-1/m)}$ is negative and therefore $\Sigma{(R=0)}$ is positive. For the remainder of this section, we specialize to a few important values of the parameters and provide closed-form expressions for the surface density. For the case of $m=2$ and an arbitrary value of $n$, the surface density is expressible in terms of elementary functions:
\begin{equation}
\Sigma{(R)} = \frac{{\Phi_0}}{Ga} \frac{n\Gamma\left(n+\frac{1}{2}\right)}{2\sqrt{\pi}\Gamma(1+n)} \frac{2+(1-2n)y^2}{{\left(1+y^2\right)}^{n+\frac{3}{2}}},
\end{equation}
where we introduced the dimensionless parameter $y = \frac{R}{a}$. The surface density for the case $m=1$ and arbitrary $n$, which includes the Hernquist model, is also analytically tractable and the final result can be expressed in terms of Gauss hypergeometric functions. We will refrain from writing down the exact expression as it is not particularly illuminating. However, for the special case of $m=1$ and $n=\frac{1}{2}$ which has an NFW-like profile, the surface density is expressible in terms of elliptic functions:
\begin{equation}
\Sigma = \frac{\Phi_0}{4\sqrt{2}{\pi}Ga}\frac{2y\left(y^2-5\right)E\left(\frac{y-1}{2y}\right)+\left(4+5y-y^3\right)K\left(\frac{y-1}{2y}\right)}{y^{1/2}\left(1-y^2\right)^{2}}.
\end{equation}
The surface densities for the $m=2$ models with differing values of $n$ are plotted for these models in Fig. \ref{figdisp2}.

\subsection{The hypervirial family}\label{sechypervirial}
The hypervirial family, whose distribution function is given by equation (\ref{DFhypervirial}), represents a special class of models which have many special features such as finite total mass, a one-term distribution function, one-term density, etc. Hence, in this section, we will list the results obtained in Section 2 for this special family of models. The potential for this family is
\begin{equation}\label{Phihypervirial}
\Phi = -\frac{\Phi_{0}}{\left(1+\left(\frac{r}{a}\right)^m\right)^{1/m}}.
\end{equation}
The density for the hypervirial family takes on a fairly simple expression
\begin{equation}
\rho(r) = \frac{\Phi_0}{4\pi{G}}\frac{(1+m)a^{-m}}{r^{2-m}\left(1+\left(\frac{r}{a}\right)^m\right)^{\frac{1}{m}+2}}.
\end{equation}
The velocity dispersion relations given by equations (\ref{sigmar}) and (\ref{sigmat}) reduce to
\begin{equation}
\sigma_r^2(r) = -\frac{1}{2}\frac{1}{1+m}\phi,
\end{equation}
\begin{equation}
\sigma_\phi^2(r) = - \frac{m}{4}\frac{1}{m+1} \phi.
\end{equation}
The radial and tangential pressures as obtained from the distribution function are given by:
\begin{equation}\label{Prhypervirial}
P_r = \frac{{\Phi_0}^2}{8{\pi}Gr^2}\left(\frac{r}{a}\right)^m \frac{1}{\left(1+\left(\frac{r}{a}\right)^m\right)^{\frac{2}{m}+2}},
\end{equation}
\begin{equation}\label{Pthypervirial}
P_\theta = P_\phi = \frac{m}{2} \frac{{\Phi_0}^2}{8{\pi}Gr^2}\left(\frac{r}{a}\right)^m \frac{1}{\left(1+\left(\frac{r}{a}\right)^m\right)^{\frac{2}{m}+2}}.
\end{equation}
It is pointed out by \citet{gf21} that these models satisfy the virial theorem $locally$, a property they term hyperviriality.

\section{A further generalization of the anisotropic polytrope models}
As we have seen, a major motivation to generalize the distribution function studied in Paper I to the family presented in this paper is the fact that the anisotropy parameter becomes position-dependent. A natural next step would be to consider a distribution function that is the sum of more than two terms, each of which is of the generalized polytrope form, i.e.
\begin{equation}\label{GD}
f = \sum_j {D_jL^{2\alpha_j}E^{\beta_j}},
\end{equation}
where all coefficients $D_{j}$'s are non-negative. Another motivation for such a distribution function is that it may be a low-energy, low-angular-momentum approximation for some other distribution function, as any distribution function can be expanded in a power series in energy and angular momentum. The corresponding Poisson equation is:
\begin{equation}\label{Poisson}
\nabla^2 \Phi = \sum_j 2^{\alpha_j+\frac{7}{2}} {\pi}^{5/2} G D_j  \Gamma\left(\alpha_j,\beta_j\right) r^{2\alpha_j} (-\Phi)^{\alpha_j + \beta_j + \frac{3}{2}}.
\end{equation}
Needless to say, directly solving such an equation is very difficult. However, it is not too hard to see that any potential of the form
\begin{equation}\label{GP}
\phi = -\Phi_0 \left({\sum_j{C_j\left(\frac{r}{a}\right)^{p_j}}}\right)^{-q},
\end{equation}
where $\Phi_{0}$ is positive, and all coefficients and powers are also positive, is consistent with a distribution function of the form (\ref{GD}). Indeed, the corresponding mass-density profile will take the same form as equation (\ref{Poisson}), allowing us to solve for the powers and coefficients in the distribution function in terms of those appearing in the potential. A potential of the form (\ref{GP}) also has many desirable features. It is the unique solution of the Poisson equation with the following set of boundary conditions: it is finite at the centre and goes to zero at infinity. Moreover, it is also a monotonic function provided all coefficients and powers are non-negative. Moreover, the requirement of non-negative coefficients and powers will also ensure that the coefficients appearing in the Poisson equation are non-negative.

On taking the Laplacian of the above potential, one notes that there are multiple ways of casting it in the same form as (\ref{Poisson}). This in turn implies that one can generate several different distribution functions of the form (\ref{GD}) that give rise to the same potential. To illustrate this point, we could arbitrarily multiply and divide equation (26) by a factor of $(1+(\frac{r}{a})^{m})$ and cast the equation as
\begin{eqnarray}
\rho &=& \left(\frac{mn\Phi_0}{4\pi G a^{2}}\right) \bigg[(m+1)\left(\frac{r}{a}\right)^{m-2} + (2+m-mn)\left(\frac{r}{a}\right)^{2m-2} \nonumber \\
&+& (1-mn)\left(\frac{r}{a}\right)^{3m-2} \bigg] \left(-\frac{\Phi}{\Phi_0}\right)^{1+\frac{3}{n}},
\end{eqnarray}
from which we can derive a three-term, two-parameter distribution function for our potential--density pair:
\begin{eqnarray}
f &=& D_{1} L^{m-2}\mathcal{E}^{\frac{1}{2}-\frac{m}{2}+\frac{3}{n}} + D_{2} L^{2m-2}\mathcal{E}^{\frac{1}{2}-m+\frac{3}{n}} \nonumber \\
&+& D_{3} L^{3m-2}\mathcal{E}^{\frac{1}{2}-\frac{3m}{2}+\frac{3}{n}},
\end{eqnarray}
where $D_{1}$, $D_{2}$ and $D_{3}$ can be written as a function of $m$ and $n$. An advantage of using a three-term distribution function compared to a two-term one is that the anisotropy profile can have more complex behaviours.

We end by noting that \citet{d29} derive a class of distribution functions that form a two-term, one-parameter family which happens to be a particular case of the family presented in this section. The distribution function and the potential density obtained by \citet{d29} are:
\begin{equation}\label{2termEvansAn}
f(\mathcal{E},L) = C_1 L^{p-2} \mathcal{E}^{3p/2+1/2} + C_2 L^{p/2-2} \mathcal{E}^{3p/4+1/2},
\end{equation}
\begin{equation}\label{PhiEvansAn}
\Phi = -\frac{GM}{\left(r^p + 2C{r_0}^{p/2}r^{p/2}+{r_0}^p\right)^{1/p}},
\end{equation}
where the coefficients $C_{1}$ and $C_{2}$ are related to $C$ by:
\begin{equation}
C_{1} = \frac{1-C^{2}}{2^{1+p/2}(2\pi)^{5/2}}\frac{\Gamma{(2p+3)}}{\Gamma{(p/2)}\Gamma{(\frac{3p}{2}+\frac{3}{2})}},
\end{equation}
\begin{equation}
C_{2} = \frac{C}{2^{1+p/4}(2\pi)^{5/2}}\frac{\Gamma{(p+3)}}{\Gamma{(p/4)}\Gamma{(\frac{3p}{4}+\frac{3}{2})}}.
\end{equation}
We note that this model is also a generalization of the hypervirial family: upon setting $C=0$ we recover equations (\ref{DFhypervirial}) and (\ref{Phihypervirial}). Moreover, this family of models also possesses the hypervirial property and therefore we will refer to it as the two-component hypervirial family. This family is studied in some detail in Appendix A.

\section{A general relativistic extension of the potential--density pair}
Given the importance of the Newtonian polytropes in modelling the structure of stars and star systems, it is natural to search for generalizations of those models in the context of general relativity. To this day, we have a handful of exact solutions which may be interpreted as relativistic versions of Newtonian polytropes, among which are the Schwarzschild constant density interior solution, a relativistic Plummer model due to \cite{b700}) and an exact solution also due to Buchdahl which generalizes the polytrope with index 1 (\cite{b81}). These solutions belong to the much broader class of exact solutions known as the static, spherically symmetric perfect fluids (SSSPF). Extensive reviews on this subject can be found at \cite{dl98} and \cite{fs}, and recent breakthroughs in the form of powerful techniques to generate SSSPF solutions can be found at \cite{l03}, \cite{mv04} and \cite{bvw05}.

\begin{figure*}
$$
\begin{array}{ccc}
 \includegraphics[width=5.28cm]{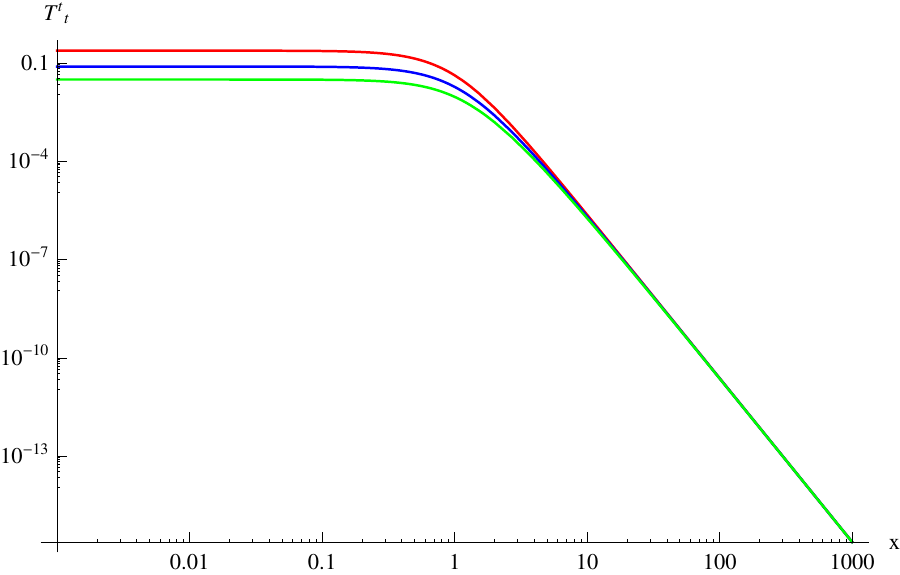} & \includegraphics[width=5.28cm]{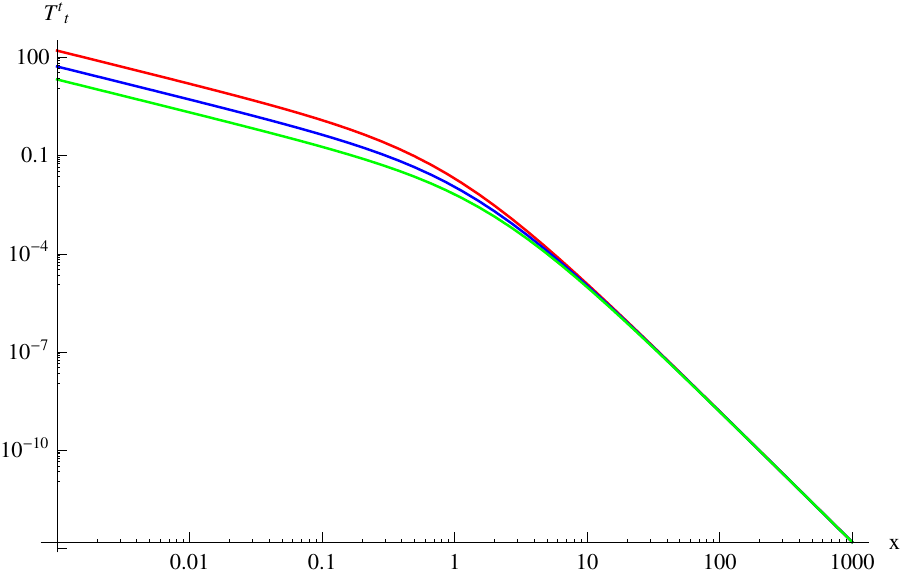} &   \includegraphics[width=5.28cm]{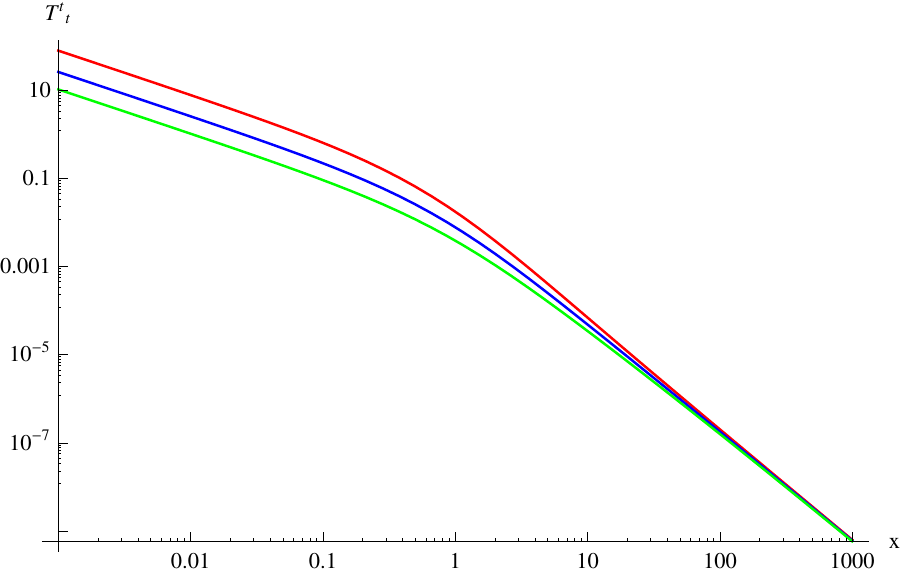}\\
 \quad\quad(a) & \quad\quad(b) &  \quad\quad(c)\\
\includegraphics[width=5.28cm]{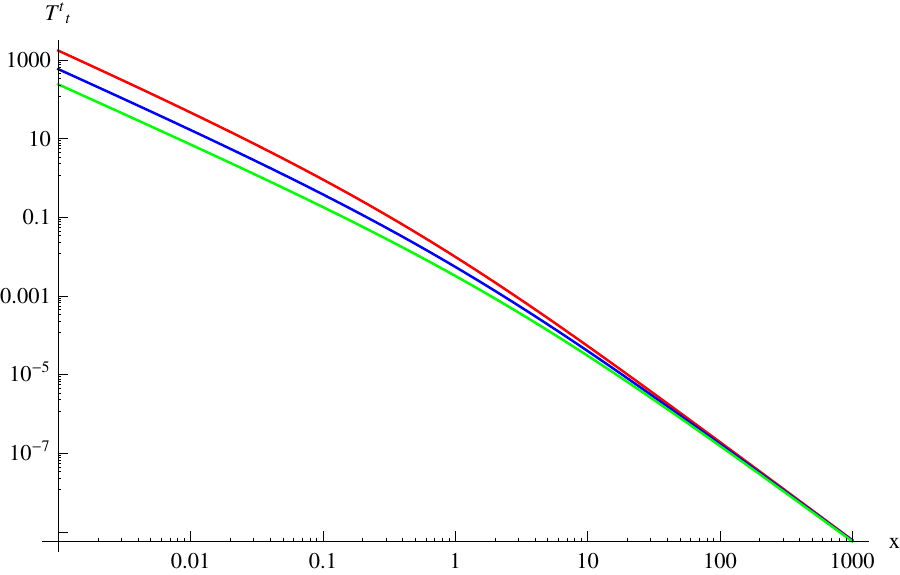} & \includegraphics[width=5.28cm]{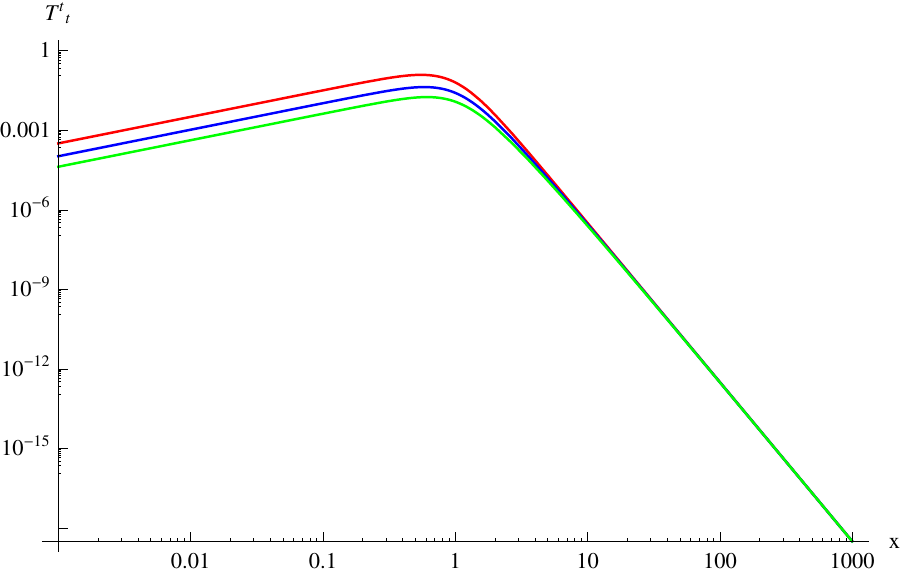} & \includegraphics[width=5.28cm]{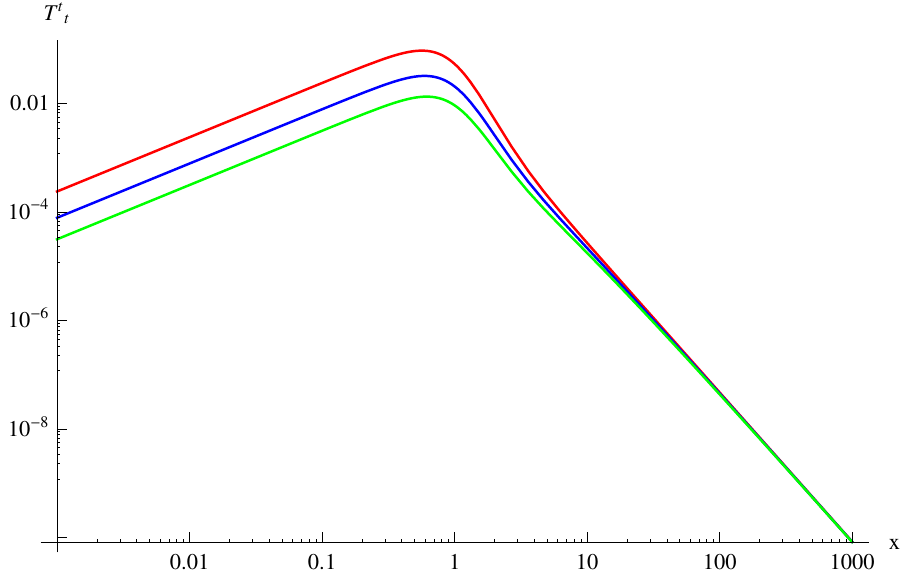}\\
 \quad\quad(d) & \quad\quad(e) & \quad\quad(f) \\
\end{array}
$$
\caption{(color figures online) The $x$-axis is expressed in units of $x=\frac{r}{a}$, while the $y$-axis is expressed in terms of the dimensionless energy density of $\tilde{T}^t_t=\frac{T^t_t\,Ga^2}{{\Phi_0}c^2}$. Figure (a) represents the Plummer model which has an inner cusp with zero slope. Figure (b) represents the Hernquist model. Figures (c) and (d) are plotted for the NFW-like profile with $m=1$, $n=\frac{1}{2}$ and for another cuspy model with $m=\frac{1}{2}$, $n=1$. Figures (e) and (f) are models that give rise to shells, and their parameters are respectively given by $m=3$, $n=\frac{1}{3}$ (a hypervirial model) and $m=3$, $n=\frac{1}{4}$.  The last two figures give rise to a different behaviour near the origin that arises from the fact that they model shells and not cusps. The green, blue and red curves represent the curves with $\lambda=1,1/2,0$, respectively. Note that this is a log-log plot.}
\label{figdisp3}
\end{figure*}

\begin{figure*}
$$
\begin{array}{ccc}
 \includegraphics[width=5.28cm]{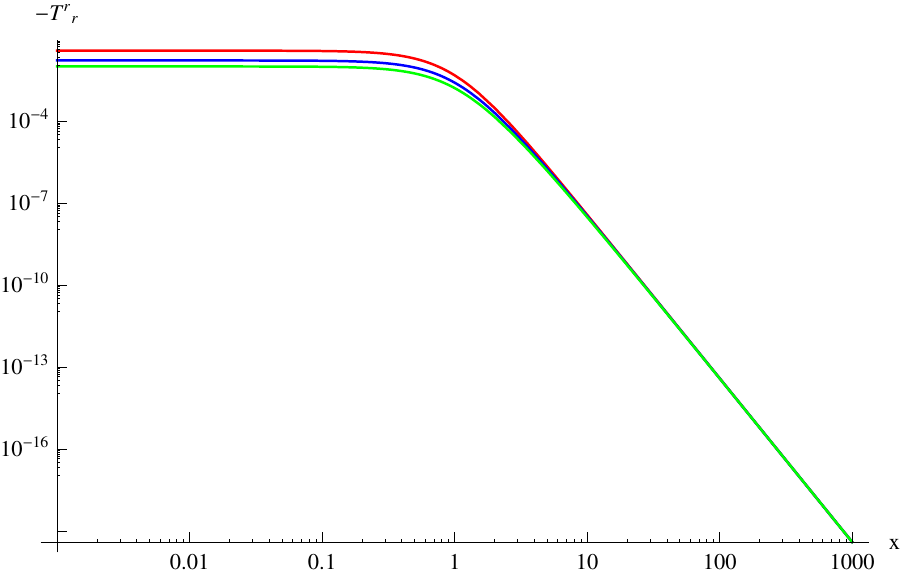} & \includegraphics[width=5.28cm]{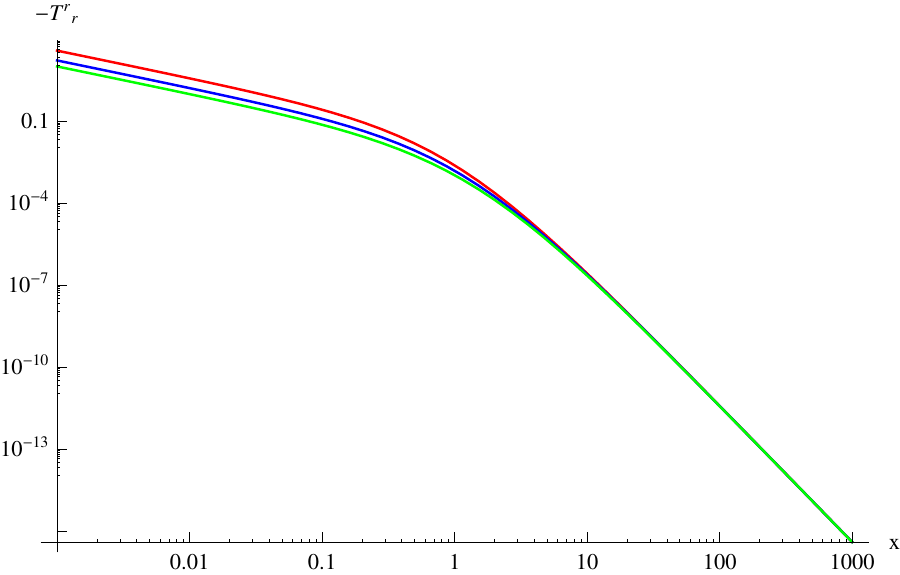} &   \includegraphics[width=5.28cm]{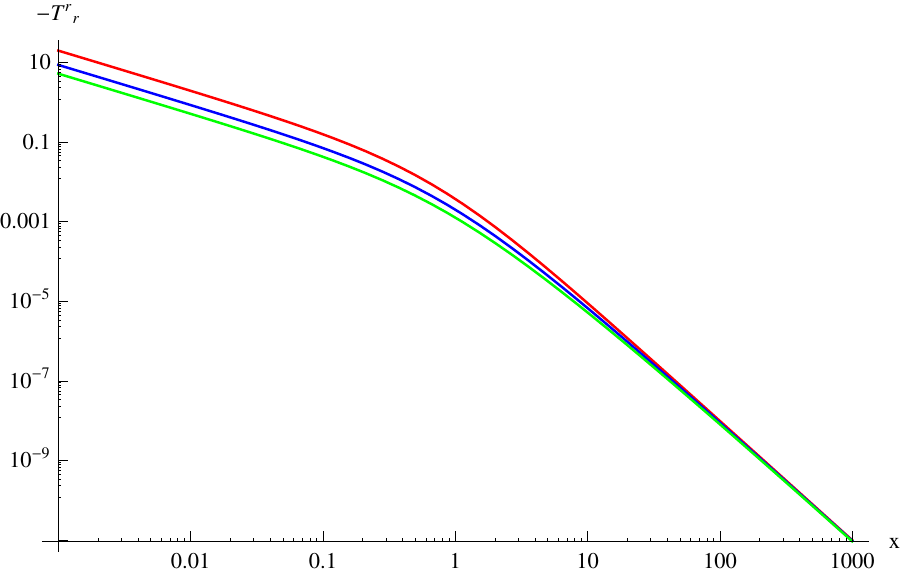}\\
 \quad\quad(a) & \quad\quad(b) &  \quad\quad(c)\\
\includegraphics[width=5.28cm]{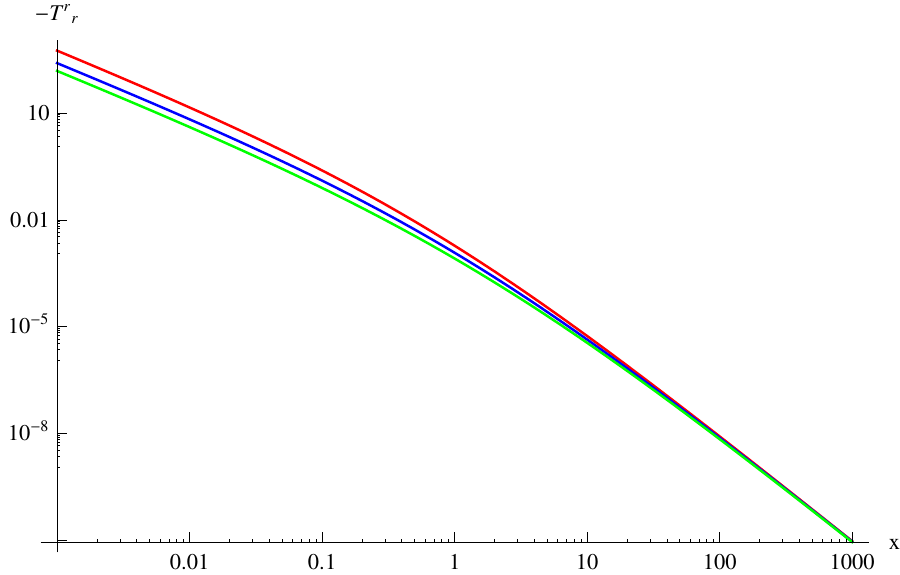} & \includegraphics[width=5.28cm]{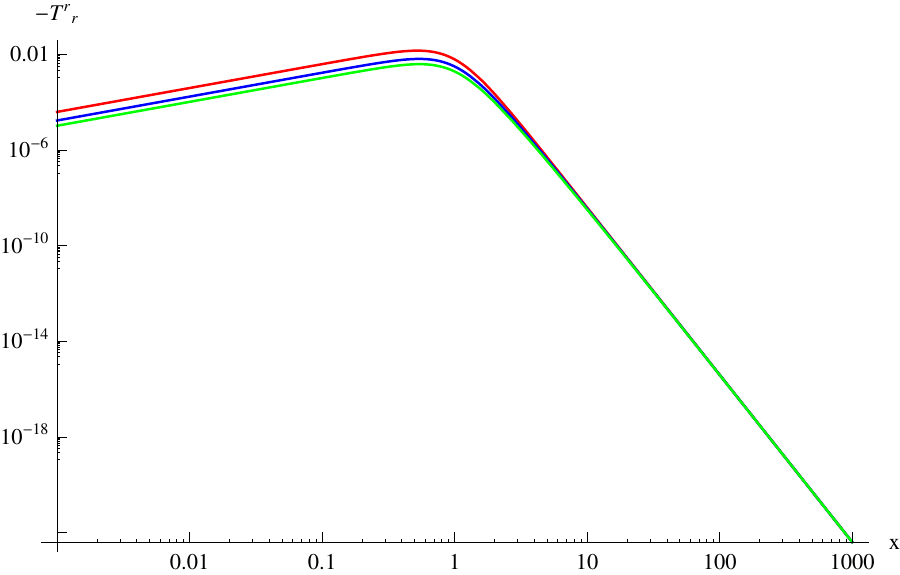} & \includegraphics[width=5.28cm]{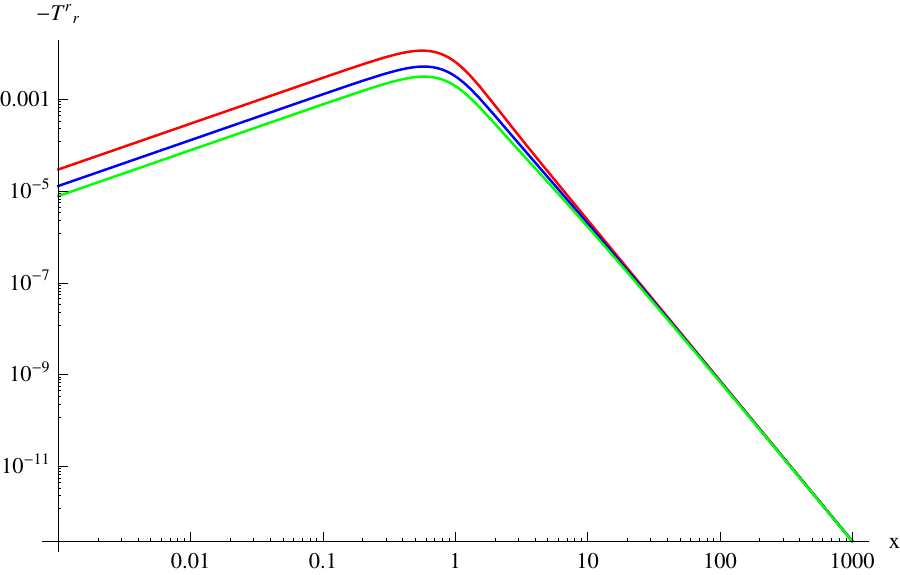}\\
 \quad\quad(d) & \quad\quad(e) & \quad\quad(f) \\
\end{array}
$$
\caption{(color figures online) The $x$-axis is expressed in units of $x=\frac{r}{a}$, while the $y$-axis is expressed in terms of the dimensionless energy density of $-\tilde{T}^r_r=-\frac{T^r_r\,Ga^2}{{\Phi_0}c^2}$. Figure (a) represents the Plummer model which has an inner cusp with zero slope. Figure (b) represents the Hernquist model. Figures (c) and (d) are plotted for the NFW-like profile with $m=1$, $n=\frac{1}{2}$ and for another cuspy model with $m=\frac{1}{2}$, $n=1$. Figures (e) and (f) are models that give rise to shells, and their parameters are respectively given by $m=3$, $n=\frac{1}{3}$ (a hypervirial model) and $m=3$, $n=\frac{1}{4}$.  The last two figures give rise to a different behaviour near the origin that arises from the fact that they model shells and not cusps. The green, blue and red curves represent the curves with $\lambda=1,1/2,0$, respectively. Note that this is a log-log plot.}
\label{figdisp4}
\end{figure*}

\begin{figure*}
$$
\begin{array}{ccc}
 \includegraphics[width=5.28cm]{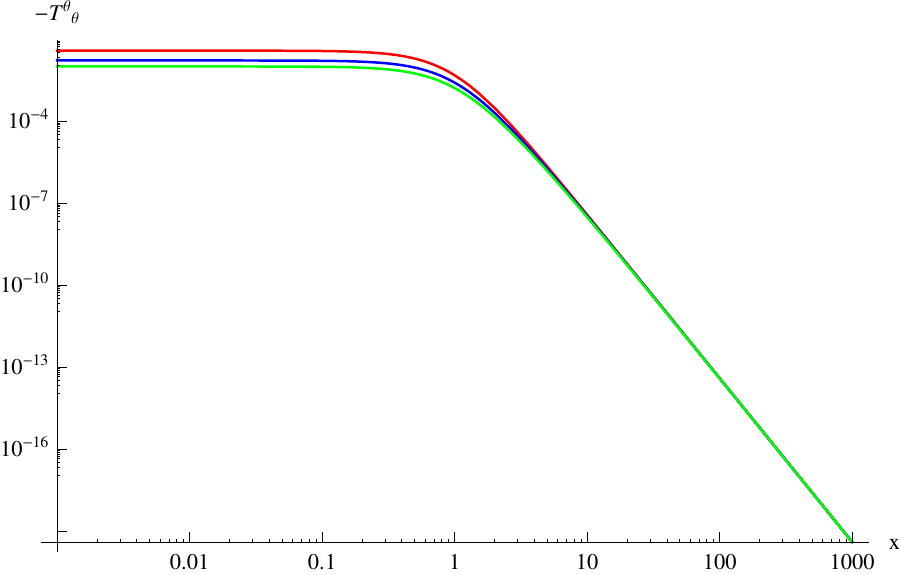} & \includegraphics[width=5.28cm]{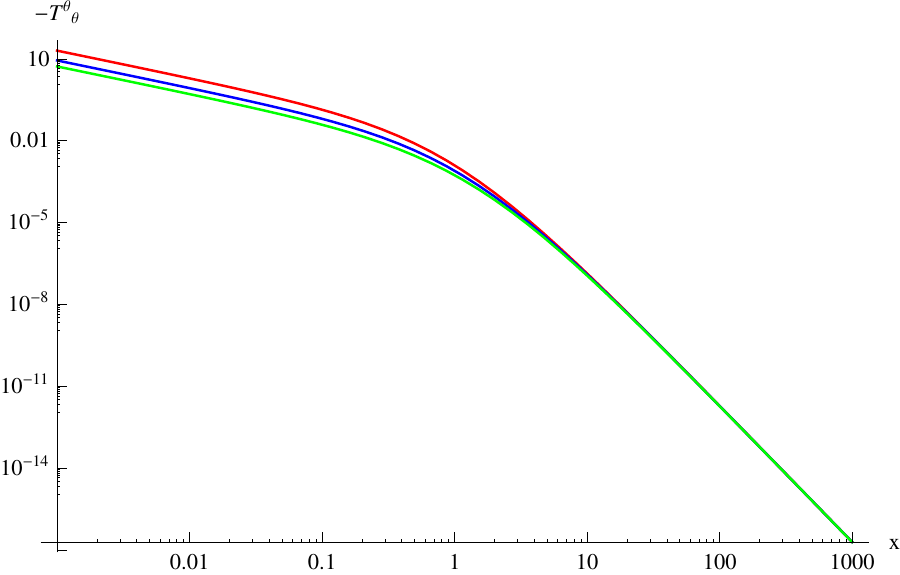} &   \includegraphics[width=5.28cm]{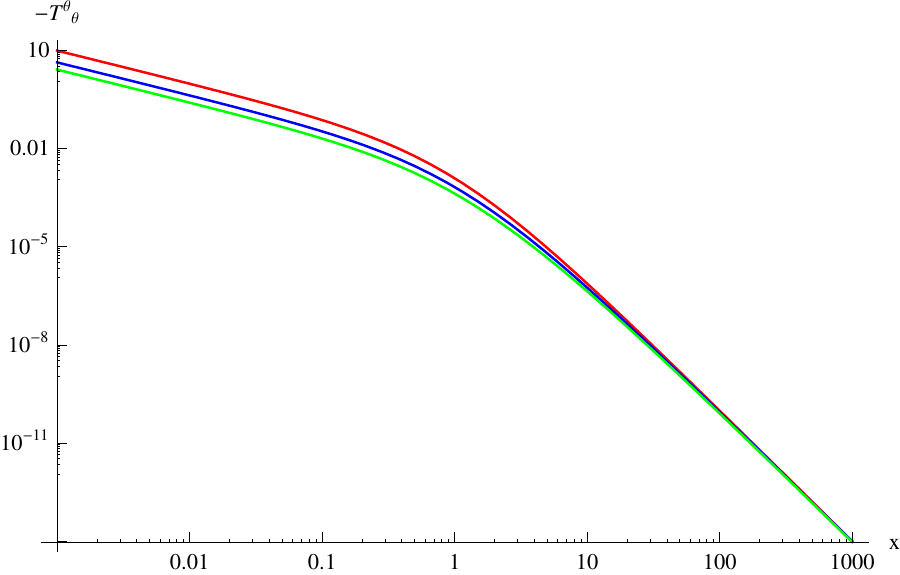}\\
 \quad\quad(a) & \quad\quad(b) &  \quad\quad(c)\\
\includegraphics[width=5.28cm]{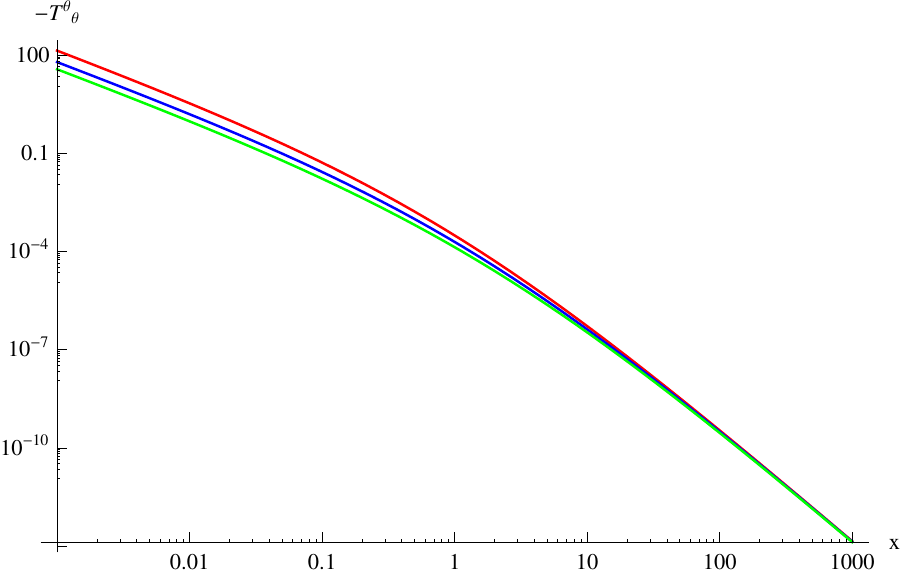} & \includegraphics[width=5.28cm]{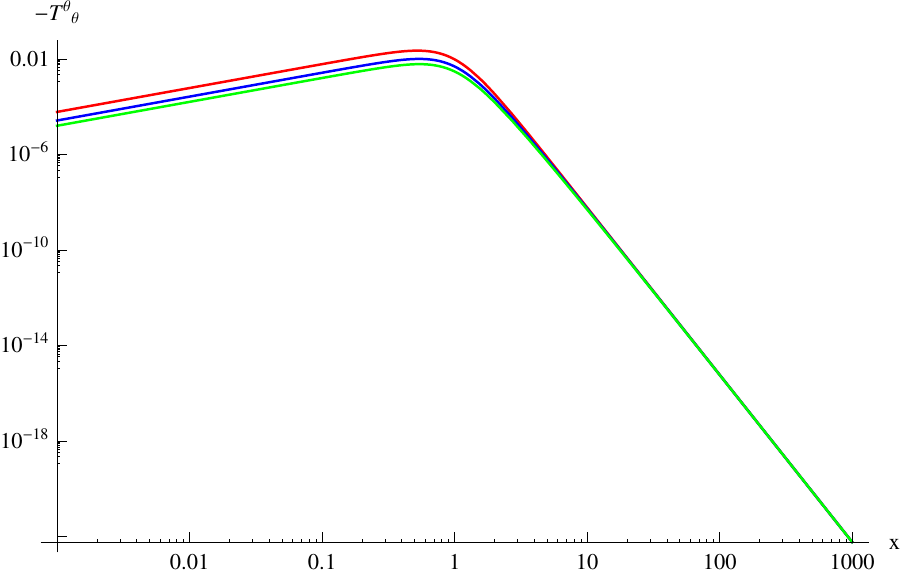} & \includegraphics[width=5.28cm]{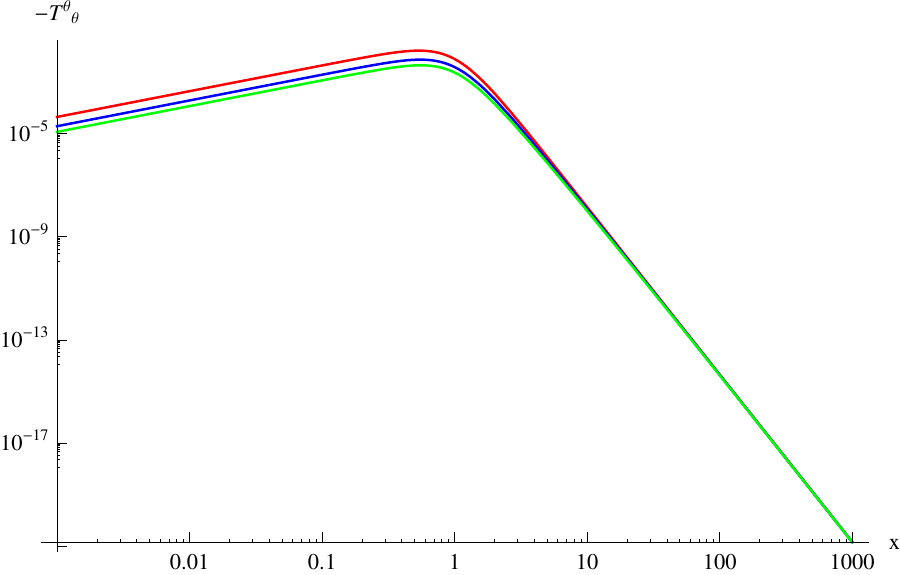}\\
 \quad\quad(d) & \quad\quad(e) & \quad\quad(f) \\
\end{array}
$$
\caption{(color figures online) The $x$-axis is expressed in units of $x=\frac{r}{a}$, while the $y$-axis is expressed in terms of the dimensionless energy density of $-\tilde{T}^\theta_\theta=-\frac{T^\theta_\theta\,Ga^2}{{\Phi_0}c^2}$. Figure (a) represents the Plummer model which has an inner cusp with zero slope. Figure (b) represents the Hernquist model. Figures (c) and (d) are plotted for the NFW-like profile with $m=1$, $n=\frac{1}{2}$ and for another cuspy model with $m=\frac{1}{2}$, $n=1$. Figures (e) and (f) are models that give rise to shells, and their parameters are respectively given by $m=3$, $n=\frac{1}{3}$ (a hypervirial model) and $m=3$, $n=\frac{1}{4}$.  The last two figures give rise to a different behaviour near the origin that arises from the fact that they model shells, and not cusps. The green, blue and red curves represent the curves with $\lambda=1,1/2,0$, respectively. Note that this is a log-log plot.}
\label{figdisp5}
\end{figure*}

Even though SSSPF solutions are intended to model stellar structures, generalizations of such solutions to describe stellar systems have also been found. For example, a dynamical analogue of Buchdahl's relativistic Plummer sphere was found by \cite{f71}. A conceptual novelty when we extend SSSPF solutions to stellar systems is the possibility of having local anisotropy. For a selection of anisotropic analogues of SSSPF solutions, see \cite{hdof01}, \cite{hmo02} and \cite{b24}. It was recently noticed by \cite{b94,b95,b96} that the approach used by Buchdahl to derive the relativistic Plummer solution can be extended to other Newtonian potentials to yield anisotropic solutions. These authors applied the approach to generate exact solutions whose Newtonian limit reduces to well-known axisymmetric and spherically symmetric galactic models, such as the Miyamoto-Nagai potential, the Satoh potential as well as the hypervirial family (only the shell-like models of this family were considered). This will be the approach taken here, and we start by writing down a generalized version of the Schwarzschild metric in isotropic coordinates:
\begin{equation}\label{isotropic}
ds^2 = \left({\frac{1-f}{1+f}}\right)^2 c^2 dt^2 - (1+f)^4 \left({dr^2+r^2 d{\theta}^2 + r^2 {\sin\theta}^2 d{\phi}^2}\right),
\end{equation}
where $f$ is a function of $r$ only. Here, we used the (+,--,--,--) signature for the sake of continuity with the existing literature on the subject. On substituting the components of the metric into the Einstein field equations, we find that the metric above is an exact solution with the following stress--energy tensor:
\begin{equation}
T^t_t = - \frac{c^4}{2{\pi}G(1+f)^5}\frac{1}{r^2}\frac{d}{dr}\left({r^2\frac{df}{dr}}\right),
\end{equation}
\begin{equation}
T_r^r = \frac{c^4}{2{\pi}G(1+f)^5(1-f)}\frac{df}{dr}\left({\frac{f}{r}+\frac{df}{dr}}\right),
\end{equation}
\begin{equation}
T_\theta^\theta = T_\phi^\phi = \frac{c^4}{4{\pi}G(1+f)^5(1-f)}\left[{f\frac{d^2f}{dr^2}+\frac{f}{r}\frac{df}{dr}-\left({\frac{df}{dr}}\right)^2}\right].
\end{equation}
The above choice of the metric is motivated by the fact that upon setting $f=\frac{GM}{2 c^2 r}$, we obtain the Schwarzschild metric in isotropic coordinates and that $\Phi=-\frac{GM}{r}$ is the Newtonian limit of the Schwarzschild metric. By extension, one can construct $f$ in such a manner that it gives rise to any arbitrary potential in the Newtonian limit. This is done by choosing $f$ to be
\begin{equation}
f = - \frac{\Phi}{2c^2}.
\end{equation}
A considerable advantage of this approach is that the Newtonian limit is readily identified, making it easy to compare with Newtonian results. Indeed, the slow-motion limit for $T^t_t$ coincides with the Laplacian of $\Phi$, which implies that the time-time Einstein field equation reduces to the Poisson equation in the Newtonian limit. To write down the components of the stress--energy tensor, we introduce the following dimensionless variables:

\begin{equation}
x = \frac{r}{a},\quad\quad \lambda = \frac{\Phi_0}{c^2},
\end{equation}
and
\begin{equation}
\xi = 1+x^{m}.
\end{equation}
The stress--energy tensor is then:
\begin{equation}
T^t_t = \frac{8{\Phi_0}{c^2}mnx^m\xi^{-2+4n}\left[1+m+(1-mn)x^m\right]}{{\pi}Ga^2x^2\left(\lambda+2\xi^n\right)^5},
\end{equation}
\begin{equation}
T^r_r = \frac{8{\Phi_0}^2mnx^m\xi^{-2+4n}\left[-1+(mn-1)x^m\right]}{{\pi} Ga^2x^2(-\lambda+2\xi^n)(\lambda+2\xi^n)^5},
\end{equation}
\begin{equation}
T_\theta^\theta = T_\phi^\phi = \frac{-4{\Phi_0}^2m^2nx^m \xi^{-2+4n}}{{\pi}Ga^2x^2(-\lambda+2\xi^n)(\lambda+2\xi^n)^5}.
\end{equation}
The various components of the stress--energy tensor are plotted for several models with different values of $\lambda$ in Figs \ref{figdisp3}, \ref{figdisp4} and \ref{figdisp5}. We also define an effective Newtonian mass density by summing up the relativistic energy density and the relativistic pressures, since they all contribute to the gravitational field:
\begin{equation}
T = T^t_t - T^r_r - T^\theta_\theta - T^\phi_\phi,
\end{equation}
\begin{equation}
T = \frac{16{\Phi_0}c^{2}mn{x^m}{\xi^{5n-2}}\left[1+m+(1-mn)x^m\right]}{{\pi}Ga^2x^2 (-\lambda+2\xi^n) (\lambda+2\xi^n)^5}.
\end{equation}
We note that, as in the Newtonian theory, only the models with $m=2$ have a finite, non-zero central energy density and pressures (except when $\lambda=2$, but we shall see in the following section that this is not allowed as per the energy conditions). Next, the various components of the stress--energy tensor and the effective Newtonian mass density are listed for the case where $mn=1$:
\begin{equation}
T^t_t = \frac{8(m+1){\Phi_0}c^2 x^m\xi^{-2+\frac{4}{m}}}{{\pi}Ga^2x^2(\lambda+2\xi^{1/m})^5},
\end{equation}
\begin{equation}
T^r_r = -\frac{8{\Phi_0}^2x^m \xi^{-2+\frac{4}{m}}}{{\pi}Ga^2x^2 (-\lambda+2\xi^{1/m})(\lambda+2\xi^{1/m})^5},
\end{equation}
\begin{equation}
T_\theta^\theta = T_\phi^\phi = \frac{m}{2}T^r_r,
\end{equation}
\begin{equation}
T = \frac{16(m+1){\Phi_0}c^2{x^m}\xi^{-2+\frac{5}{m}}}{{\pi}Ga^2x^2(-\lambda+2\xi)^{1/m})(\lambda+2\xi^{1/m})^5}.
\end{equation}

\subsection{Energy conditions}
We follow the approach outlined in \citet{b95} to obtain the criteria for the various forms of the energy condition which are discussed in this section.

In this section, the case $mn \leq 1$ is investigated, and we discover that the condition $\lambda <1 $ ensures that the energy conditions are satisfied for all values of $r$. The case $mn \geq 1$ can be similarly undertaken, but the energy conditions will only hold for a finite range of $r$. The cutoff radius is exactly the same as the Newtonian cutoff radius for the density, given by equation (\ref{cutoffr}).

The $weak$ energy condition states that 
\begin{equation}
T^t_t \geq 0.
\end{equation}
Since $T^t_t$ is the energy density as seen by a stationary observer, the above condition is equivalent to the condition that the energy density is positive in a said frame. By starting with the expression for $T^t_t$, the weak energy condition reduces to the constraint that $mn \leq 1$ is satisfied everywhere for all values of $r$. We point out the fact that this condition is exactly identical to the one obtained in the Newtonian case by imposing the condition that $\rho \geq 0$. 

The strong energy condition states that one requires $T \geq 0$. This reduces to the condition that
\begin{equation}
\frac{(1+m)+(1-mn)x^m}{-\lambda+2\left(1+x^m\right)^n} \geq 0.
\end{equation}
We note that the numerator is always positive, regardless of the value of $m$ and $n$ in the range of interest. Hence, the inequality purely depends on the sign of the denominator. The term $\left(1+x^m\right)^n$ is always greater than unity, and choosing 
\begin{equation}
\lambda < 2
\end{equation}
ensures that the denominator is always positive, regardless of the value of $x$.

The $dominant$ energy condition states that the momentum flux in any given frame can never be superluminal. Algebraically speaking, this is equivalent to the set of conditions
\begin{equation}
\left|\frac{T^r_r}{T^t_t}\right| \leq 1, \quad \left|\frac{T^\theta_\theta}{T^t_t}\right| \leq 1, \quad \left|\frac{T^\phi_\phi}{T^t_t}\right| \leq 1.
\end{equation}
On substituting the relevant expressions, one obtains that
\begin{equation}\label{relineq1}
\left|\frac{\lambda\left[-1+(mn-1)x^m\right]}{\left[(1+m)+(1-mn)x^m\right]\left[-\lambda+2\left(1+x^m\right)^n\right]}\right| \leq 1,
\end{equation}
\begin{equation}\label{relineq2}
\left|\frac{\lambda{m}}{2\left[(1+m)+(1-mn)x^m\right]\left[-\lambda+2\left(1+x^m\right)^n\right]}\right| \leq 1.
\end{equation}
Let us consider the first inequality above for the case $mn \leq 1$. Notice that we can split up the left-hand side into the product:
\begin{equation}
\left|\frac{1+(1-mn)x^{m}}{1+m+(1-mn)x^{m}}\right|\left|\frac{\lambda}{\lambda-2(1+x^{m})^{n}}\right| \leq 1.
\end{equation}
Since $1-mn \geq 0$ and $m \geq 0$, the first factor in equation (\ref{relineq1}) has a magnitude that is at most 1. Thus, if we can show that the second factor also has magnitude at most 1 (for some range of $\lambda$), then the inequality above holds for that range of $\lambda$. We then have:
\begin{equation}
\lambda \leq \left|\lambda - 2(1+x^{m})^{n}\right|.
\end{equation}
But by the strong energy condition, $\lambda < 2$ so the quantity under the absolute value sign on the right-hand side is negative. Therefore
\begin{equation}
\lambda \leq 2(1+x^{m})^{n} - \lambda.
\end{equation}
This inequality is satisfied for all $x$ if $\lambda < 1$.

Let us consider the second inequality, given by equation (\ref{relineq2}), arising from the dominant energy condition. We specialize to the case where $mn \leq 1$ and use the condition that we obtained from the first energy condition, i.e. $\lambda < 1$. The second condition can be rewritten as follows:
\begin{equation}
\left|\frac{\frac{1}{2}m}{1+m+(1-mn)x^{m}}\right|\left|\frac{\lambda}{\lambda-2(1+x^{m})^{n}}\right| \leq 1.
\end{equation}
With the above constraints, the second factor becomes lesser than unity for all values of $x$. The first factor reduces to the inequality
\begin{equation}
\left|\frac{\frac{1}{2}m}{1+m}\right| < 1.
\end{equation}
But, this inequality is always satisfied for $m >0$.

\subsection{Comments on the pressure}
The approach described above is rather ad hoc: even though the stress--energy tensor obtained by substituting the metric into Einstein's equation satisfies the energy conditions, there is no particular physical basis for it. In order to gain understanding of the stress--energy tensor, we want to compute the Newtonian limit for the pressure. Expanding $T_r^r$, $T^\phi_\phi$ and $T^\theta_\theta$ to lowest order in $\lambda$, we find:
\begin{equation}
P_r = -T^r_r = \frac{{\Phi_0}^2mnx^m\left[1+(1-mn)x^m\right]}{8{\pi}Ga^2x^2\left(1+x^m\right)^{2+2n}},
\end{equation}
\begin{equation}
P_\theta = P_\phi = -T_\theta^\theta = - T_\phi^\phi = \frac{{\Phi_0}^2m^2nx^m}{16{\pi}Ga^2x^2\left(1+x^m\right)^{2+2n}}.
\end{equation}
From these two expressions, we see that the Newtonian limit for the pressures does not coincide with the Newtonian pressures computed from the distribution function (equations \ref{Pr} and \ref{Pt}). This to be expected, as this approach only guarantees that the correct expressions for the potential--density pair are recovered in the slow-moving limit (expansion in powers of $c^{-1}$) but not the pressure.

Remarkably, it is found that the Newtonian limit for the pressure $does$ coincide with the expressions derived earlier for the special case of $mn=1$, i.e. equations (\ref{Prhypervirial}) and (\ref{Pthypervirial}). Thus, in this particular case the stress--energy tensor can be justified on dynamical grounds, at least in the Newtonian regime. This leads us to suppose that there exists a relativistic distribution function which gives rise to the line element (\ref{isotropic}) and whose Newtonian limit has the generalized polytrope form (\ref{DFhypervirial}). Such a relativistic distribution function probably does not have the polytropic form however.

Next, we note that the anisotropy parameter (\ref{anisotropyparameter}) can be expressed in terms of the Newtonian pressures as
\begin{equation}\label{relativisticbeta}
\beta = 1 - \frac{P_{\theta}}{P_{r}}.
\end{equation}
We may define a relativistic anisotropy parameter by replacing the Newtonian pressures by their relativistic counterparts in the expression above. For our family of models, this gives
\begin{equation}
\beta = 1 - \frac{m}{2}\left(\frac{1}{1+(1-mn)x^{m}}\right),
\end{equation}
which of course is different from equation (\ref{anisotropyparameter}). Remarkably, however, for the hypervirial case, the relativistic anisotropy parameter is exactly the same as the Newtonian one (\ref{anisotropyparameterhypervirial}).

\subsection{A study of circular orbits}

\begin{figure*}
$$
\begin{array}{ccc}
 \includegraphics[width=5.28cm]{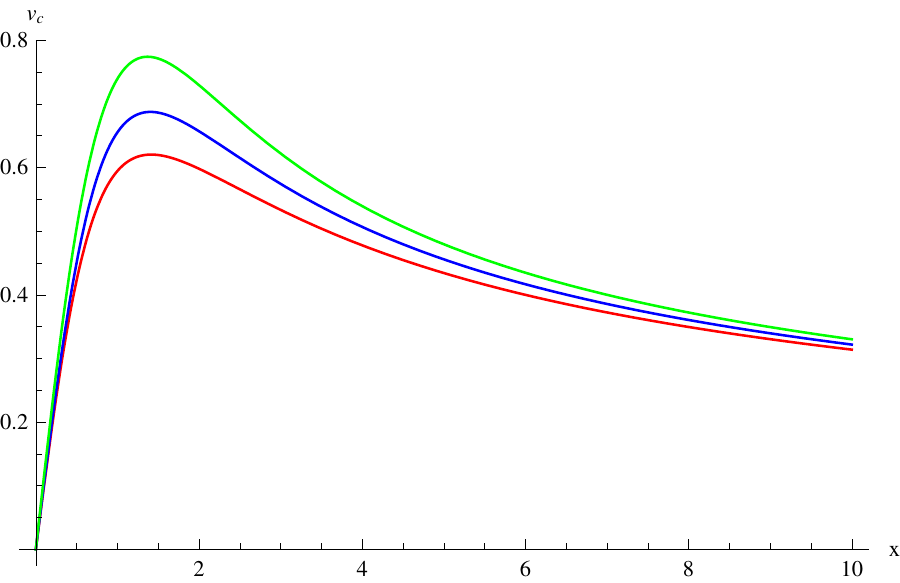} & \includegraphics[width=5.28cm]{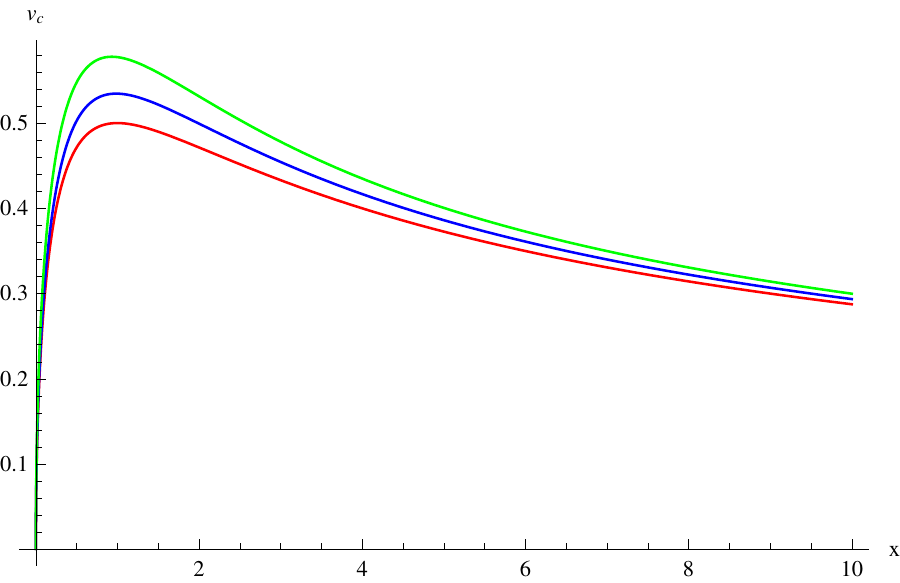} &   \includegraphics[width=5.28cm]{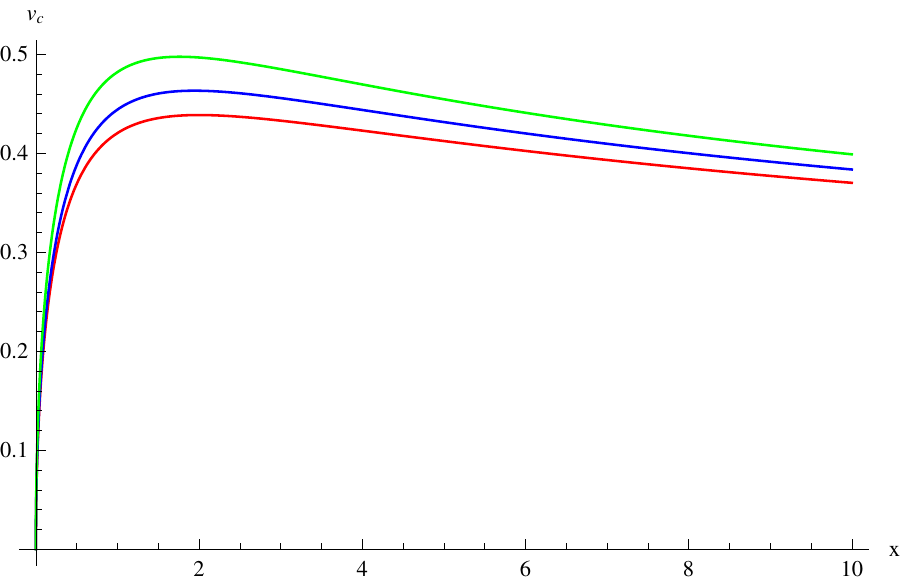}\\
 \quad\quad(a) & \quad\quad(b) &  \quad\quad(c)\\
\includegraphics[width=5.28cm]{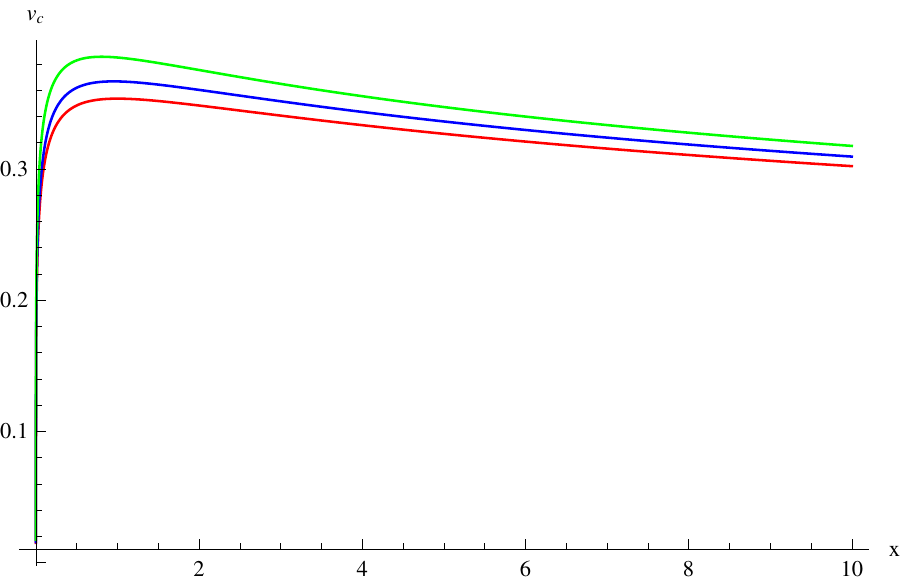} & \includegraphics[width=5.28cm]{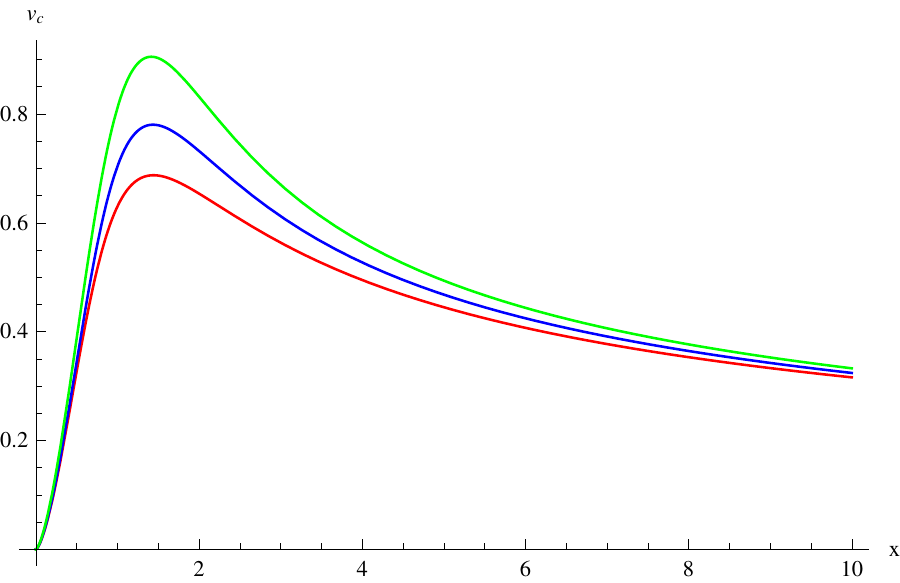} & \includegraphics[width=5.28cm]{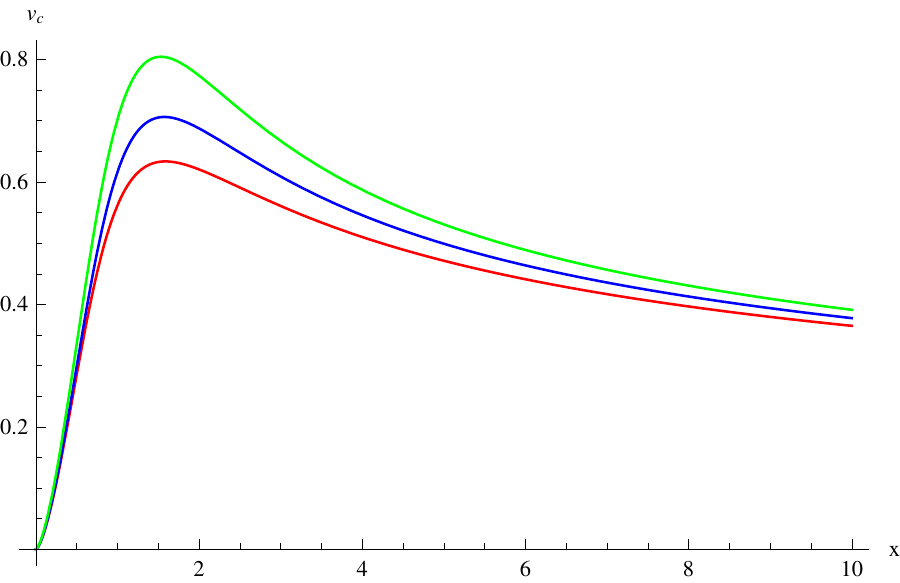}\\
 \quad\quad(d) & \quad\quad(e) & \quad\quad(f) \\
\end{array}
$$
\caption{(color figures online) The $x$-axis is expressed in units of $x=\frac{r}{a}$, while the $y$-axis is expressed in terms of the dimensionless velocity $\tilde{v}_c=\frac{v_c}{{\Phi_0}}$. Figure (a) represents the Plummer model which has an inner cusp with zero slope. Figure (b) represents the Hernquist model. Figures (c) and (d) are plotted for the NFW-like profile with $m=1$, $n=\frac{1}{2}$ and for another cuspy model with $m=\frac{1}{2}$, $n=1$. Figures (e) and (f) are models that give rise to shells, and their parameters are, respectively, given by $m=3$, $n=\frac{1}{3}$ (a hypervirial model) and $m=3$, $n=\frac{1}{4}$. The green, blue and red curves represent the curves with $\lambda=1,1/2,0$, respectively. }
\label{figdisp6}
\end{figure*}

\begin{figure*}
$$
\begin{array}{ccc}
 \includegraphics[width=5.28cm]{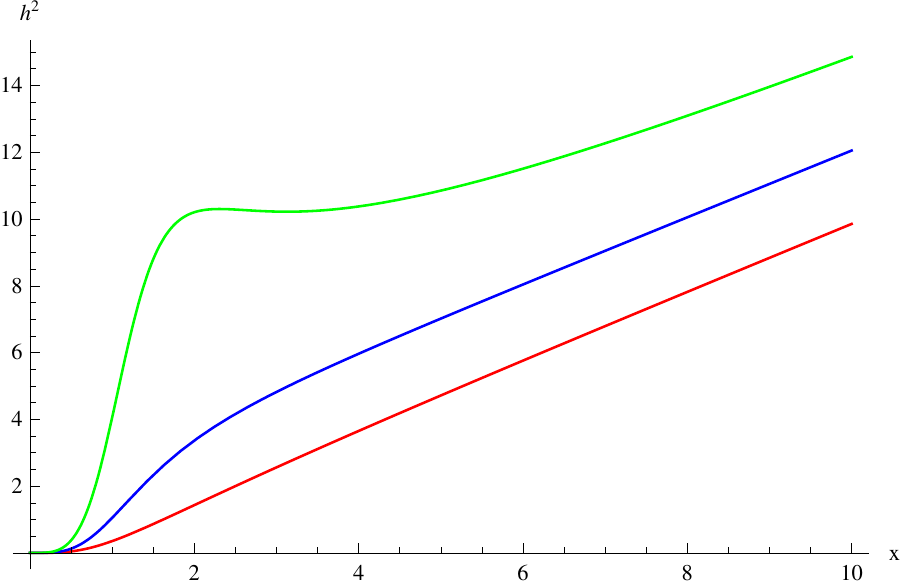} & \includegraphics[width=5.28cm]{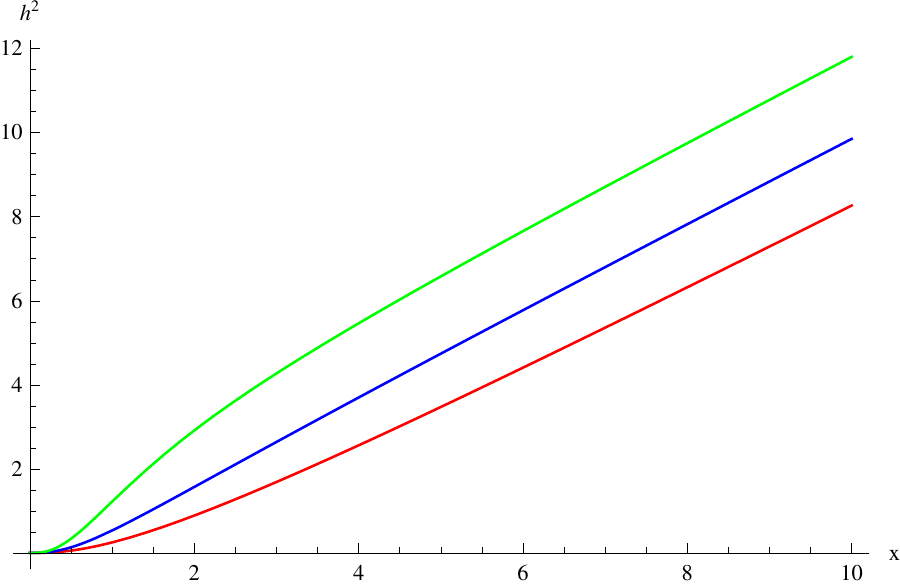} &   \includegraphics[width=5.28cm]{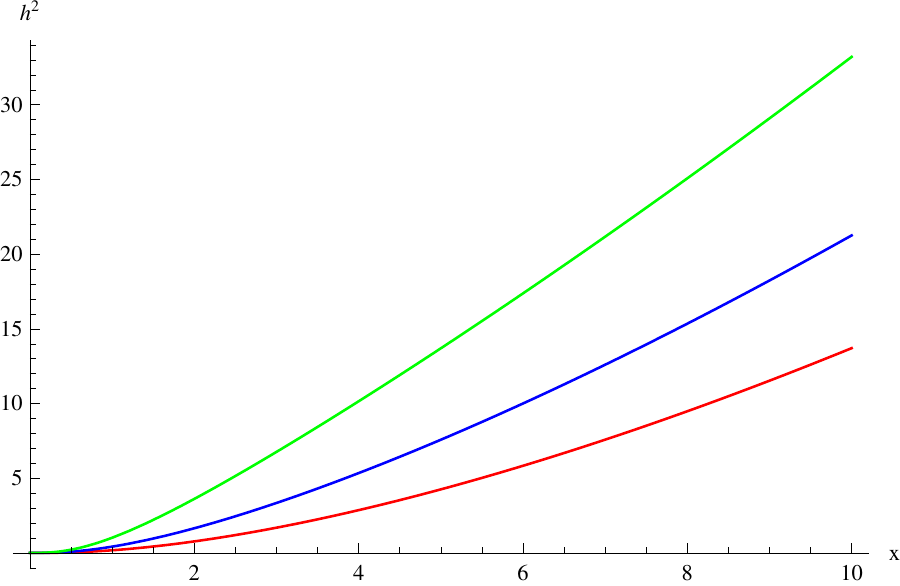}\\
 \quad\quad(a) & \quad\quad(b) &  \quad\quad(c)\\
\includegraphics[width=5.28cm]{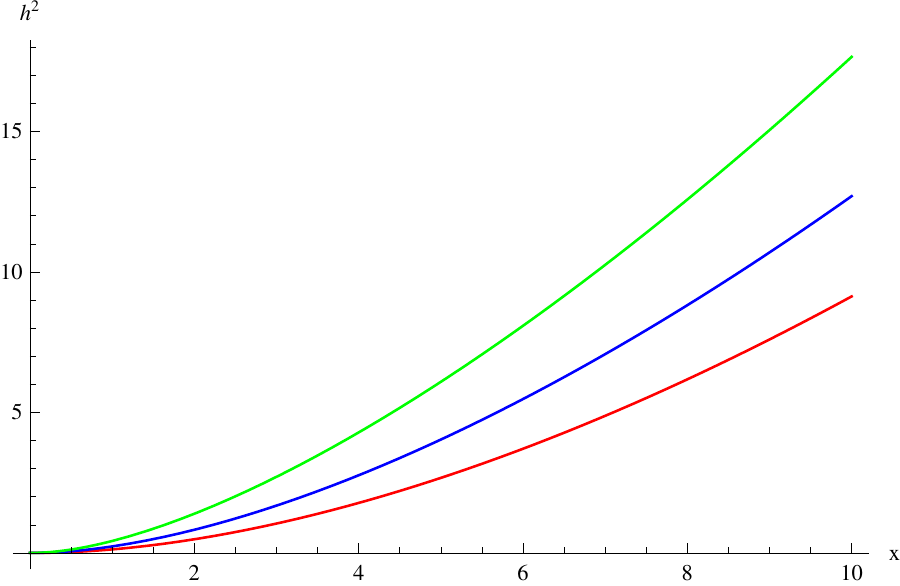} & \includegraphics[width=5.28cm]{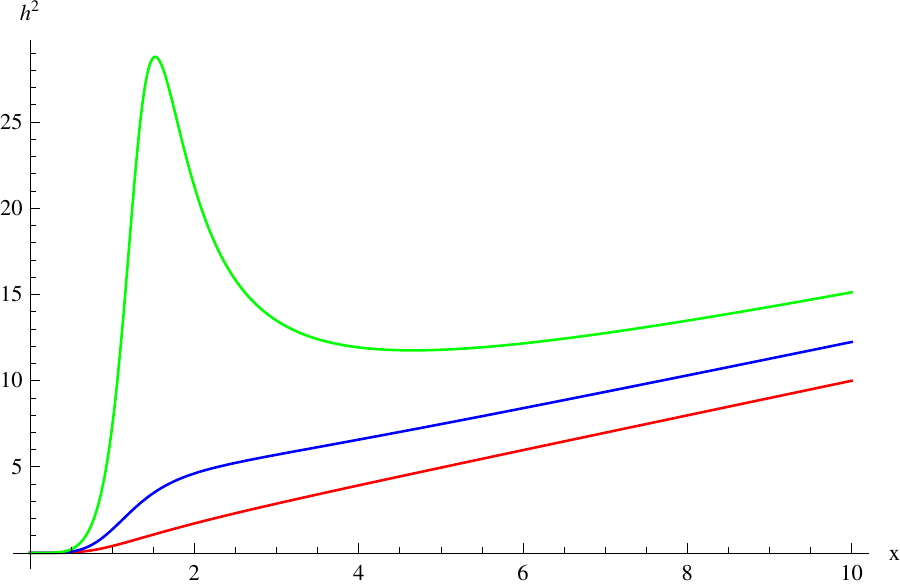} & \includegraphics[width=5.28cm]{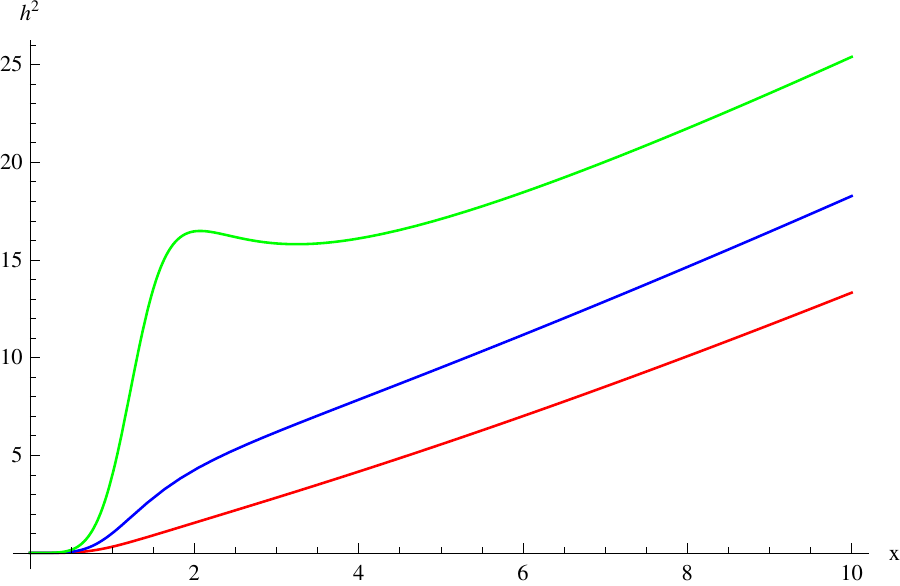}\\
 \quad\quad(d) & \quad\quad(e) & \quad\quad(f) \\
\end{array}
$$
\caption{(color figures online) The $x$-axis is expressed in units of $x=\frac{r}{a}$, while the $y$-axis is expressed in terms of the dimensionless angular momentum squared which is given by $\tilde{h}^2=\frac{h^2}{{r^2}{\Phi_0}}$. Figure (a) represents the Plummer model which has an inner cusp with zero slope. Figure (b) represents the Hernquist model. Figures (c) and (d) are plotted for the NFW-like profile with $m=1$, $n=\frac{1}{2}$ and for another cuspy model with $m=\frac{1}{2}$, $n=1$. Figures (e) and (f) are models that give rise to shells, and their parameters are, respectively, given by $m=3$, $n=\frac{1}{3}$ (a hypervirial model) and $m=3$, $n=\frac{1}{4}$. The green, blue and red curves represent the curves with $\lambda=1,1/2,0$, respectively. }
\label{figdisp7}
\end{figure*}

The relativistic effects on the rotation profiles of galaxies are a somewhat controversial topic. Even though most astrophysical systems fall under the slow-moving regime and so should be well described by Newtonian mechanics, it is obviously important that we should be able to analyse galactic dynamics using general relativity. Moreover, radical proposals have been made that relativistic corrections can satisfactorily explain the discrepancy between the Newtonian rotation profiles and the observed ones \citep*[see][]{b11,b26}, thereby rendering the dark matter hypothesis unnecessary.

According to these authors, the analysis of gravitationally bound systems such as galaxies or star clusters is an intrinsically non-linear problem for which Newtonian gravity (which is the linearization of general relativity) is ill-suited. Because such non-linearities should reveal themselves at the first post-Newtonian order, an analysis based on relativistic kinetic theory was carried out in Paper I to this order of approximation. It is found in that paper that post-Newtonian effects can potentially produce a flatter rotation curve compared to the Newtonian theory.

Consider a test particle moving in a circular orbit of radius $r$ in a spacetime described by equation (\ref{isotropic}). The velocity of the particle as measured from an observer at infinity can be computed from the geodesic equation, and is given by (\cite{b95})
\begin{equation}
v_c = \sqrt{\frac{-2c^2rf_{,r}}{(1-f)\left(1+f+2rf_{,r}\right)}}.
\end{equation}
For our choice of $f$, we find
\begin{equation}
v_c = \sqrt{\frac{4{\Phi_0}mnx^m \xi^n}{(-\lambda+2\xi^n)[\lambda-\lambda(2mn-1)x^m+2\xi^{n+1}]}}.
\end{equation}
Taking the Newtonian limit, the above expression coincides with the result obtained by the theory of central potentials in classical mechanics, with the circular velocity given by
\begin{equation}
v_c = \sqrt{r\frac{d\Phi}{dr}} = \sqrt{\frac{{\Phi_0}mn\,x^m}{\left(1+x^m\right)^{n+1}}}.
\end{equation}
The relativistic circular velocity falls off at large distance as
\begin{equation}
v_{c} \propto x^{-mn/2}.
\end{equation}
Note that this is independent of $\lambda$, i.e. the large $r$ behaviour is the same as in the Newtonian theory. In particular, we have the Keplerian result $v_{c} \propto x^{-1/2}$ for the hypervirial family, and by choosing $mn \approx 0$, we can obtain arbitrarily flat rotation curves. 

Fig. \ref{figdisp6} depicts the dimensionless rotation curves as a function of $x$. We note that the relativistic curves are slightly steeper than the Newtonian ones, but they exhibit the same behaviour qualitatively. In particular, Figs \ref{figdisp6}(c) and (d) have relatively flat rotation curves, which match well with their NFW-like nature. It is also worth mentioning that in both Paper I and \cite{b11}, new degrees of freedom appear in the relativistic model when compared to the Newtonian model. In Paper I, for instance, the central value of the post-Newtonian field $\psi$ becomes freely adjustable, whereas in \cite{b11}, the solution is an infinite series where each term comes with a freely specifiable coefficient. Our present approach only allows for one free parameter ($\lambda$, which determines how relativistic the system is) and therefore gives a more limited range of scenarios.

For completeness, we give the expression for the relativistic circular velocity for the case of $mn=1$ as
\begin{equation}
v_c = \sqrt{\frac{4{\Phi_0}x^m\xi^{1/m}}{(-\lambda+2\xi^{1/m})[\lambda(1-x^m)+2\xi^{\frac{1}{m}+1}]}},
\end{equation}
and the Newtonian limit is given by
\begin{equation}
v_c = \sqrt{r\frac{d\Phi}{dr}} = \sqrt{\frac{{\Phi_0}\,x^m}{\left(1+x^m\right)^{\frac{1}{m}+1}}}.
\end{equation}

Finally, we study the specific angular momentum of test particles in circular orbits. This is given by (\cite{b95})
\begin{equation}
h = cr^2(1+f)^2\sqrt{\frac{-2f_{,r}}{r\left[1-f^2+2rf_{,r}(2-f)\right]}}.
\end{equation}
For our choice of $f$, we find
\begin{equation}
h = \frac{a\sqrt{\Phi_{0}mn}x^{1+\frac{m}{2}}\xi^{-2n}(\lambda+2\xi^{n})^{2}}{2\sqrt{-8\lambda mnx^{m}+4\xi^{n+1}+\lambda^{2}\xi^{-n}[-1+(2mn-1)x^{m}]}}.
\end{equation}
In the Newtonian limit, this reduces to the classical expression
\begin{equation}
h = r v_c = ax \sqrt{\frac{{\Phi_0}mn\,x^m}{\left(1+x^m\right)^{n+1}}}.
\end{equation}
For the special case of $mn=1$, we obtain
\begin{equation}
h = \frac{a\sqrt{\Phi_{0}}x^{1+\frac{m}{2}}\xi^{-\frac{2}{m}}(\lambda+2\xi^{1/m})^{2}}{2\sqrt{-8\lambda x^{m}+4\xi^{1+\frac{1}{m}}+\lambda^{2}\xi^{-1/m}(-1+x^{m})}},
\end{equation}
and the Newtonian limit for this special case is
\begin{equation}
h = r v_c = a x \sqrt{\frac{{\Phi_0}\,x^m}{\left(1+x^m\right)^{\frac{1}{m}+1}}}.
\end{equation}
One can use Rayleigh's criterion to investigate the stability of circular orbits. The condition for stability is (\cite{b95})
\begin{equation}
\frac{d\left(h^2\right)}{dr} \geq 0.
\end{equation}
Fig. \ref{figdisp7} depicts the dimensionless angular momentum as a function of the scaled distance. From the plots, it can be seen that the addition of relativity results in a fairly significant distortion for some of the profiles. The shell-like profiles, represented by Figs \ref{figdisp7}(e) and (f), are seen to exhibit a region of negative slope. This violates Rayleigh's criterion and indicates that the circular orbits are unstable in this region. Notice that this instability is a purely relativistic effect, and the fact that shell configurations are subject to this effect is corroborated by Paper I, \cite{b95} as well as numerical studies such as \cite{a07}.

\subsection{Comments on the total mass}
It is pointed out by \cite{gf21} that the hypervirial family has a few special properties, among which is the fact that they have finite total mass despite their infinite extent. We show in this section that the finiteness of the total mass still holds in the relativistic case. The proper total mass is given by
\begin{equation}
Mc^{2} = \int_{0}^{\infty} \sqrt{-h} T_{t}^{t} 4\pi r^{2} dr,
\end{equation}
where $\sqrt{-h}$ is the determinant of the spatial metric:
\begin{equation}
\sqrt{-h} = \left(1 + f{(r)}\right)^6.
\end{equation}
The total mass in terms of $f{(r)}$ is then
\begin{equation}
M = -\frac{2c^{2}}{G} \int_{0}^{\infty} (1+f{(r)}) \frac{d}{dr}\left(r^{2}\frac{df}{dr}\right) dr.
\end{equation}
The particular case $m=2$ is described in \citet{b700}. For general $m$, we find
\begin{equation}
M = \frac{\Phi_{0}a}{G} + \frac{\Phi_{0}^{2}a}{2Gc^{2}} \frac{\Gamma{(2+\frac{1}{m})}\Gamma{(1+\frac{1}{m})}}{\Gamma{(2+\frac{2}{m})}},
\end{equation}
where the lowest order can be recognized as the Newtonian total mass, and the second term is the relativistic correction. Since $\Phi_{0}$ is bounded from above by the energy conditions, the total mass of the system is also bounded from above. By making use of the strong energy condition, we obtain the following upper bound:
\begin{equation}
\frac{GM}{ac^{2}} < 2 + 2 \frac{\Gamma{(2+\frac{1}{m})}\Gamma{(1+\frac{1}{m})}}{\Gamma{(2+\frac{2}{m})}}.
\end{equation}
\section{Conclusions}

In this paper, we have presented an extension of the generalized polytropes that were derived in \citet{gf21} and Paper I. The extended model is constructed by means of the superposition of two such terms and gives rise to a simple two-component, two-parameter model. On solving for the potential--density pair, we find that it includes the hypervirial models (of which the Plummer and Hernquist models are special cases), a family of models similar to the Ossipkov--Merrit models, and a density profile that closely approximates the NFW profile. The velocity dispersion relations are obtained, and the anisotropy parameter is found to be non-constant, except for the hypervirial family of models. We also impose several constraints on the range of $r$ and our parameters by demanding that they maintain physical and mathematical consistency. On computing the projected quantities, we discover that analytical expressions can be obtained for a subclass of these models. Anisotropic polytropes, owing to their simplicity, are shown to be easily generalized to an even wider class of distribution functions, which in turn give rise to a larger family of potential--density pairs. In Appendix B, we investigate in greater detail the models that resemble the Ossipkov--Merritt family, and show that they possess an isotropic core and a radially anisotropic outer region, which is of relevance for dark matter haloes (e.g. \citet{hm06}).

In the second section, we obtain the relativistic extension of the above potential--density pair by following an approach along the lines of \citet{b700} and \citet{b94,b95}. The relativistic pressure is obtained through the stress--energy tensor, and its Newtonian limit is derived. This is compared against the pressure calculated in the first section, and the two results are found to be different. This is to be expected since this approach only guarantees that the correct potential--density pairs are recovered in the Newtonian limit and not necessarily the pressure. However, for the case of the hypervirial family, we discover that the two expressions for the pressure coincide, suggesting that there exists a relativistic distribution function for this family of models, which reduces to the generalized polytropic distribution function discussed here and in Paper I. Furthermore, we also discover that the relativistic and the Newtonian anisotropic parameters exactly coincide for the hypervirial family. In Appendix A, we also show that the two anisotropic parameters coincide for a different hypervirial family derived by \citet{d29}.

The strong, dominant and weak energy conditions are studied to obtain a permitted range of values for the parameters and $r$. We find that the value of $mn=1$ is a critical value that divides the parameter space into two different regimes. The relativistic rotation curves are obtained and shown to reduce to the familiar Newtonian expressions in the slow-moving limit. These curves are found to be steeper than their Newtonian counterparts, but they possess the same qualitative behaviour. The relativistic angular momentum is also computed, and it is also used, via Rayleigh's criterion, to investigate orbital stability. We discover that relativistic effects are solely responsible for orbital instability, which also depends on the choice of the free parameters. For the relativistic hypervirial family, we obtain an upper bound for the mass of the system. 

The methods used in this paper, involving anisotropic polytropes and the relativistic counterparts for the potential--density pairs, allow one to tackle a broad range of spherically symmetric systems. In the future, we hope to use extensions of these methods to study other astrophysical structures that cannot be directly modeled by using the techniques outlined in this paper.

\section*{Acknowledgments}
We are grateful to Richard Matzner, Philip Morrison, Juan Pedraza, Tanja Rindler-Daller and Lawrence Shepley for their illuminating discussions and comments. The research was supported in part by NASA under grant no. NNX09AU86G.

\appendix
\section{The relativistic generalization of the two-component hypervirial potential}
We apply the anisotropic analogue of the SSSPF as outlined in Section 2. We compute the various components of the stress--energy tensor
and express the end result in terms of the following dimensionless variables:
\begin{equation}
\lambda = \frac{GM}{bc^2},\quad x = \frac{r}{r_0}, \quad \xi = \left(1+x^p+2Cx^{p/2}\right).
\end{equation}
From equation (\ref{PhiEvansAn}), we find the following components of the stress--energy tensor and effective Newtonian density:
\begin{equation}
T^t_t = \frac{Mc^2}{4{\pi}b^3}\frac{\left[C(2+p)+2\left(1+C^2+p\right)x^{p/2}+C(2+p)x^p\right]}{\xi^{2-4/p}x^{2-p/2}\left[\lambda+\xi^{1/p}\right]^5},
\end{equation}
\begin{equation}
T_r^r = \frac{GM^2}{2{\pi}b^4}\frac{\left[1+Cx^{p/2}\right]\left[C+x^{p/2}\right]}{\xi^{2-4/p}x^{2-p/2}\left[\lambda+\xi^{1/p}\right]^5\left[-\lambda+\xi^{1/p}\right]},
\end{equation}
\begin{equation}
T_\phi^\phi = T^\theta_\theta = \frac{GM^2}{8{\pi}b^4}\frac{p\left[C+2x^{p/2}+Cx^{p}\right]}{\xi^{2-4/p}x^{2-p/2}\left[\lambda+\xi^{1/p}\right]^5\left[-\lambda+\xi^{1/p}\right]},
\end{equation}
\begin{equation}
T = \frac{Mc^2}{4{\pi}b^3}\frac{\left[C(2+p)+2\left(1+C^2+p\right)x^{p/2}+C(2+p)x^p\right]}{\xi^{2-5/p}x^{2-p/2}\left[\lambda+\xi^{1/p}\right]^5\left[-\lambda+\xi^{1/p}\right]}.
\end{equation}
One can obtain a bound for the value of $\lambda$ by making use of the strong energy condition. We note that \cite{d29} point out the allowed range of values for $C$ and $p$, namely that the former lies between $0$ and $1$, and that the latter is always positive. Through the strong energy condition, this amounts to merely requiring that the denominator is positive. This condition is satisfied provided that the condition  $\lambda < 1$ is satisfied. This coincides exactly with our result obtained for our earlier potential, despite the fact that the potentials are quite dissimilar.

Finally, we give the relativistic anisotropy parameter as defined in equation (\ref{relativisticbeta}):
\begin{equation}
\beta = 1 - \frac{p}{4}\left[\frac{C+2x^{p/2}+Cx^{p}}{(1+Cx^{p/2})(C+x^{p/2})}\right].
\end{equation}
Rather remarkably, this is exactly the same as the Newtonian anisotropy parameter given in \cite{d29}.

\section{An Ossipkov--Merritt-like family of distribution functions}\label{appOssipkovMerritt}
In this appendix, we will introduce a specific family of distribution functions that are closely associated with the Ossipkov--Merritt distribution functions. Let us consider equation (\ref{DF0}) and substitute $m=2$ in it. The distribution function reduces to
\begin{equation} \label{DF101}
f = B \mathcal{E}^{ \frac{2}{n} - \frac{3}{2}} \left[ \mathcal{E} + \frac{1}{2}\left(\frac{2}{n}-\frac{1}{2}\right) a^{-2}\left(\frac{1-2n}{3}\right) L^{2} \right] .
\end{equation}
Let us now suppose that $n$ takes on the form
\begin{equation}
n = \frac{4}{3+2\zeta},\quad\quad\quad \zeta = 0,1,2, \ldots
\end{equation}
For these values of $n$, $\frac{2}{n} - \frac{3}{2} = \zeta$, and hence, $\mathcal{E}$ is raised to integral powers. For $\zeta \geq 3$, the value of $mn \leq 1$, and one can define the anisotropy radius
\begin{equation}
r_a = a \sqrt{\frac{3(3+2\zeta)}{(1+\zeta)(2\zeta-5)}} ,
\end{equation}
and $Q_+$ is the usual Ossipkov--Merritt variable, given by
\begin{equation}
Q_+ =  \mathcal{E} + \frac{L^2}{2r_a^2} .
\end{equation}
Thus, the distribution function will become
\begin{equation}
f = B \left(Q_+ - \frac{L^2}{2r_a^2}\right)^{\zeta} Q_+ ,
\end{equation}
and since $\zeta$ is an integer, the distribution function reduces to
\begin{equation}
f = B \sum_{k=0}^\zeta \frac{\zeta{!}}{(\zeta-k){!}k{!}} {Q_+}^{k+1} L^{2\zeta-2k}  (-1)^{\zeta-k} \left(\frac{1}{2r_a^2}\right)^{\zeta-k} .
\end{equation}
Now, if one considers the values of $\zeta = 0,1,2,\dots$ the situation is slightly different. Then, we must redefine the anisotropy radius and the variable $Q_{-}$ as follows:
\begin{equation}
r_a = a \sqrt{\frac{3(3+2\zeta)}{(1+\zeta)(5-2\zeta)}} ,
\end{equation}
\begin{equation}
Q_{-} =  \mathcal{E} - \frac{L^2}{2r_a^2} .
\end{equation}
The distribution function will be given by
\begin{equation}
f = B \left(Q_{-} + \frac{L^2}{2r_a^2}\right)^{\zeta} Q_{-} .
\end{equation}
Using the fact that $\zeta$ is an integer, the above distribution function reduces to
\begin{equation}
f = B \sum_{k=0}^\zeta \frac{\zeta{!}}{(\zeta-k){!}k{!}} {Q_{-}}^{k+1} L^{2\zeta-2k} \left(\frac{1}{2r_a^2}\right)^{\zeta-k} .
\end{equation}
In both the scenarios, we note that the distribution function is of the form
\begin{equation}
f \propto \sum_i C_i Q^{a_1} L^{2b_1} .
\end{equation}
Here, we note that the coefficients $C_i$ are all positive, which allows us to apply a similar approach to the one used in the earlier sections. In other words, the above distribution function becomes identically zero for $Q<0$ and takes on the above form for $Q>0$; this allows us to impose an upper bound on $Q$ when evaluating that integral.
This distribution function is a sum of terms involving integer powers of $Q$ and $L$. One such distribution function that is akin to the above equation has been studied by \citet{w32}, which builds on the work undertaken by \citet{b201}, \citet{b200} and \citet{cu91}.

In order to compute $\rho$ as a function of $r$ and $\Phi$, we use a similar approach to the one outlined in \citet{w32}. The density is given by
\begin{equation}
\rho\left(\Phi,r\right)=\frac{2\pi B}{r^{2}}\sum_{k=0}^{\zeta}\frac{\zeta!}{k!\left(\zeta-k\right)!}\left(\frac{1}{2r_{a}^{2}}\right)^{\zeta-k}I_{1}\times I_{2} ,
\end{equation}
where $I_1$ and $I_2$ are
\begin{equation}
I_{1}=\intop_{0}^{-\Phi}Q_{-}^{k+1}dQ_{-} ,
\end{equation}
and

\begin{equation}
I_{2}=\intop_{0}^{L_{0}}\frac{L^{2\left(\zeta-k\right)}dL^{2}}{\sqrt{2\left(-\Phi-Q_{-}\right)-\frac{L^{2}}{r^{2}}\left(1+\frac{r^{2}}{r_{a}^{2}}\right)}} .
\end{equation}
 
The upper bound $L_0$ is
\begin{equation}
L_{0}=2r^{2}\frac{\left(-\Phi-Q_{-}\right)}{\left(1+\frac{r^{2}}{r_{a}^{2}}\right)} .
\end{equation}
On simplification, this integral reduces to
\begin{equation}
\rho\left(\Phi,r\right)=B\sum_{k=0}^{\zeta}\frac{\zeta!}{k!\left(\zeta-k\right)!}\left(\frac{1}{2r_{a}^{2}}\right)^{\zeta-k}\Theta\left(r\right)\times I_{3} ,
\end{equation}
where the quantities $\Theta(r)$ and $I_3$ are  given by
\begin{equation}
\Theta\left(r\right)=2^{\zeta-k+\frac{3}{2}}\pi^{\frac{3}{2}}\frac{\Gamma\left(\zeta-k+1\right)}{\Gamma\left(\zeta-k+\frac{3}{2}\right)}r^{2\left(\zeta-k\right)}\left(1+\frac{r^{2}}{r_{a}^{2}}\right)^{-1-\zeta+k},
\end{equation}
\begin{equation}
I_{3}=\intop_{0}^{-\Phi}\left(-\Phi-Q_{-}\right)^{\zeta-k+\frac{1}{2}}Q_{-}^{k+1}dQ_{-} .
\end{equation}
On simplifying $I_3$, one obtains
\begin{equation}
I_{3}=\frac{\Gamma(2+k)\Gamma\left(\frac{3}{2}-k+\zeta\right)}{\Gamma\left(\zeta+\frac{7}{2}\right)}\left(-\Phi\right)^{\frac{5}{2}+\zeta} .
\end{equation}
Thus, the final integral is found to be
\begin{eqnarray}
\rho\left(\Phi,r\right)&=&\left(2\pi\right)^{\frac{3}{2}}B\left(-\Phi\right)^{\frac{5}{2}+\zeta}\frac{\zeta!}{\Gamma\left(\zeta+\frac{7}{2}\right)} \times \\ \nonumber
&& \sum_{k=0}^{\zeta}\left(\frac{r^{2}}{r_{a}^{2}}\right)^{\zeta-k}(k+1)\left(1+\frac{r^{2}}{r_{a}^{2}}\right)^{-1-\zeta+k} .
\end{eqnarray}
We note that the above formula is only true for $\zeta=0,1,2$. For values of $\zeta\geq3$, the value of $mn<1$ and one can resort to the results obtained in the earlier sections. However, for these three special values, the density must be separately evaluated,
\begin{equation}
\zeta=0\quad\Longrightarrow\quad\rho\left(\Phi,r\right)=2^{9/2}B\left(-\Phi\right)^{5/2}\frac{\pi}{15}\left(1+\frac{r^{2}}{r_{a}^{2}}\right)^{-1}.
\end{equation}
For the value of $\zeta=1$, the density is
\begin{equation}
\rho\left(\Phi,r\right) = B\left(-\Phi\right)^{7/2}\frac{\pi}{105}\left(2+3\frac{r^{2}}{r_{a}^{2}}\right)\left(1+\frac{r^{2}}{r_{a}^{2}}\right)^{-2} .
\end{equation}
And for the value of $\zeta=2$, we obtain
\begin{eqnarray}
\rho\left(\Phi,r\right) &=& 2^{15/2}B\left(-\Phi\right)^{9/2}\frac{\pi}{945}\left(1+\frac{r^{2}}{r_{a}^{2}}\right)^{-3} \\ \nonumber
&& \times \left[6\left(\frac{r^{2}}{r_{a}^{2}}\right)^{2}+8\left(\frac{r^{2}}{r_{a}^{2}}\right)+3\right] .
\end{eqnarray}
The case of $\zeta=0$ is the simplest case of the Ossipkov--Merritt model, and its distribution function corresponds to the following form:
\begin{equation}
f = B \left(\mathcal{E} - \frac{L^2}{2r^2_a}\right) .
\end{equation}
The anisotropy radius is given by $r_a = \frac{3}{\sqrt{5}}a$. For this model, the velocity dispersion relations are
\begin{equation}
\sigma_r^2(r) = -\frac{2}{7} \Phi
\end{equation}
and
\begin{equation}
\sigma_\phi^2(r) = -\frac{2}{7} \Phi \left[\frac{1}{\left(1+\frac{r^2}{r_a^2}\right)}\right] .
\end{equation}
The anisotropy parameter can be calculated and is
\begin{equation}\label{ap1}
\beta = \frac{\left(\frac{r^2}{r_a^2}\right)}{\left(1+\frac{r^2}{r_a^2}\right)} .
\end{equation}
For the case of $\zeta=1$, the distribution function is
\begin{equation}
f = B\mathcal{E}\left(\mathcal{E}-\frac{L^2}{5a^2}\right),
\end{equation}
and one can evaluate the velocity dispersion relations:
\begin{equation}
\sigma_r^2(r) = -\frac{2}{9} \Phi .
\end{equation}
\begin{equation}
\sigma_\phi^2(r) = -\frac{2}{9} \Phi \left[{\frac{\left(2+4\frac{r^{2}}{r_{a}^{2}}\right)}{\left(2+3\frac{r^{2}}{r_{a}^{2}}\right)\left(1+\frac{r^{2}}{r_{a}^{2}}\right)}}\right] .
\end{equation}
The anisotropy parameter is
\begin{equation}\label{ap2}
\beta = \frac{\left(\frac{r^{2}}{r_{a}^{2}}\right)\left(1+3\frac{r^{2}}{r_{a}^{2}}\right)}{\left(2+3\frac{r^{2}}{r_{a}^{2}}\right)\left(1+\frac{r^{2}}{r_{a}^{2}}\right)} .
\end{equation}
Finally, for the case of $\zeta=2$, the distribution function is
\begin{equation}
f = B \mathcal{E}^{2} \left(\mathcal{E}-\frac{L^2}{14a^2}\right) .
\end{equation}
The velocity dispersion relations are
\begin{equation}
\sigma_r^2(r) = -\frac{2}{11} \Phi ,
\end{equation}
\begin{equation}
\sigma_\phi^2(r) = -\frac{2}{11} \Phi \left[{\frac{\left(3+10\frac{r^{2}}{r_{a}^{2}}+10\frac{r^{4}}{r_{a}^{4}}\right)}{\left(3+8\frac{r^{2}}{r_{a}^{2}}+6\frac{r^{4}}{r_{a}^{4}}\right)\left(1+\frac{r^{2}}{r_{a}^{2}}\right)}}\right].
\end{equation}
The anisotropy parameter is found to be
\begin{equation}\label{ap3}
\beta = \frac{\left(\frac{r^{2}}{r_{a}^{2}}\right)\left(1+4\frac{r^{2}}{r_{a}^{2}}+6\frac{r^{4}}{r_{a}^{4}}\right)}{\left(3+8\frac{r^{2}}{r_{a}^{2}}+6\frac{r^{4}}{r_{a}^{4}}\right)\left(1+\frac{r^{2}}{r_{a}^{2}}\right)} .
\end{equation}
From the above expressions for the anisotropy parameter given by equations (\ref{ap1}), (\ref{ap2}) and (\ref{ap3}), we can see that  $\beta \approx 0$ for small values of $r$ and $\beta \approx 1$, i.e. $\beta>0$, for large values of $r$. Hence, the density profiles with $\zeta=0,1,2$ correspond to systems with isotropic cores and are radially anisotropic in the outer regions. Such profiles are believed to be of importance in modelling dark matter haloes (\citet{hm06}), although the extent of radial anisotropy exhibited by the above profiles is too large (\citet{ea06}).

\end{document}